\documentclass{aa}

\usepackage[utf8]{inputenc}
\usepackage{graphicx}
\usepackage[export]{adjustbox}
\usepackage{txfonts}
\usepackage{float}
\usepackage{xcolor}
\usepackage{natbib}
\setcitestyle{notesep={ }}
\usepackage[nottoc]{tocbibind}
\usepackage{wasysym}
\usepackage{threeparttable}
\usepackage{ulem}
\usepackage{appendix}
\usepackage{microtype}
\usepackage[colorlinks,linkcolor=blue]{hyperref}
\hypersetup{                    % parametrage des hyperliens
    colorlinks=true,            % colorise les liens
    breaklinks=true,            % permet les retours à la ligne pour les liens trop longs
    urlcolor= blue,             % couleur des hyperliens
    linkcolor= black,           % couleur des liens internes aux documents (index, figures, tableaux, equations,...)
    citecolor= blue             % couleur des liens vers les references bibliographiques
    }
\usepackage[nameinlink,capitalise]{cleveref}

\usepackage{enumitem}
\let\orgautoref\autoref
\providecommand{\Autoref}
        {\def\equationautorefname{Equation}%
         \def\figureautorefname{Figure}%
         \def\subfigureautorefname{Figure}%
         \def\sectionautorefname{Section}%
         \def\subsectionautorefname{Section}%
         \def\subsubsectionautorefname{Section}%
         \def\Itemautorefname{Item}%
         \def\tableautorefname{Table}%
         \orgautoref}

\renewcommand{\autoref}
        {\def\equationautorefname{Eq.}%
         \def\figureautorefname{Fig.}%
         \def\subfigureautorefname{Fig.}%
         \def\sectionautorefname{Sect.}%
         \def\subsectionautorefname{Sect.}%
         \def\subsubsectionautorefname{Sect.}%
         \def\Itemautorefname{item}%
         \def\tableautorefname{Table}%
         \orgautoref}

\begin{document} 

   \title{CORALIE radial-velocity search for companions\\around evolved stars (CASCADES)}

%   \subtitle{I. Sample definition and first results: Three new planets orbiting giant stars \thanks{Based on observations collected with the CORALIE echelle spectrograph on the 1.2-m Euler Swiss telescope at La Silla Observatory, ESO, Chile.}\fnmsep\thanks{The complete Table 2 is only available in electronic form at the CDS via anonymous ftp to }}
  \subtitle{I. Sample definition and first results: Three new planets orbiting giant stars \thanks{Based on observations collected with the CORALIE echelle spectrograph on the 1.2-m Euler Swiss telescope at La Silla Observatory, ESO, Chile.}\fnmsep\thanks{The complete Table 2 is only available in electronic form at the CDS via anonymous ftp to \href{http://cdsarc.u-strasbg.fr}{cdsarc.u-strasbg.fr/} (130.79.128.5) or via \href{http://cdsweb.u-strasbg.fr/cgi-bin/qcat?J/A\string+A/}{CDS}}}

   \author{G. Ottoni\inst{1}\and
          S. Udry\inst{1}\and
          D. Ségransan\inst{1}\and
          G. Buldgen\inst{1}\and
          C. Lovis\inst{1}\and
          P. Eggenberger\inst{1}\and
          C. Pezzotti\inst{1}\and
          V. Adibekyan\inst{2,3}\and\\
          M. Marmier\inst{1}\and
          M. Mayor\inst{1}\and
          N.C. Santos\inst{2,3}\and
          S.G. Sousa\inst{2}\and
          N. Lagarde\inst{4}\and
          C. Charbonnel\inst{1}
          }

    \institute{Département d'Astronomie, Université de Genève, Chemin Pegasi 51, 1290 Versoix, Switzerland. \email{gael.ottoni@unige.ch}
    \and
    Instituto de Astrof\'isica e Ci\^encias do Espa\c{c}o, Universidade do Porto, CAUP, Rua das Estrelas, 4150-762 Porto, Portugal
    \and 
    Departamento de F\'isica e Astronomia, Faculdade de Ci\^encias, Universidade do Porto, Rua do Campo Alegre, 4169-007 Porto, Portugal 
    \and
    Institut UTINAM, CNRS UMR 6213, Université Bourgogne Franche-Comté, OSU THETA Franche-Comté-Bourgogne, Observatoire de Besançon, BP 1615, 25010, Besançon Cedex, France}

   \date{Received December 7th, 2020; accepted November 29th 2021}

% \abstract{}{}{}{}{} 
% 5 {} token are mandatory
 
  \abstract
    {Following the first discovery of a planet orbiting a giant star in 2002, we started the CORALIE radial-velocity search for companions around evolved stars (CASCADES). We present the observations of three stars conducted at the 1.2\,m Leonard Euler Swiss telescope at La Silla Observatory, Chile, using the CORALIE spectrograph.}
    {We aim to detect planetary companions to intermediate-mass G- and K- type evolved stars and perform a statistical analysis of this population. We searched for new planetary systems around the stars HD\,22532 (TIC\,200851704), HD\,64121 (TIC\,264770836), and HD\,69123 (TIC\,146264536).}
    {We have followed a volume-limited sample of 641 red giants since 2006 to obtain high-precision radial-velocity measurements. We used the Data \& Analysis Center for Exoplanets (DACE) platform to perform a radial-velocity analysis to search for periodic signals in the line profile and activity indices, to distinguish between planetary-induced radial-velocity variations and stellar photospheric jitter, and to search for significant signals in the radial-velocity time series to fit a corresponding Keplerian model.}
    {In this paper, we present the survey in detail, and we report on the discovery of the first three planets of the sample around the giant stars HD\,22532, HD\,64121, and HD\,69123.}
  {}


   \maketitle

%-------------------------------------------------------------------
\section{Introduction}\label{sec:intro}

Since the discovery of 51 Peg\,b by \citet{Mayor1995}, the first extrasolar planet orbiting a solar-like star, over 4800 exoplanets\footnote{See e.g.,  \href{https://exoplanetarchive.ipac.caltech.edu/}{https://exoplanetarchive.ipac.caltech.edu} (as of January 5, 2022).} have been detected, including almost 900 using the radial-velocity technique. Those distant worlds cover a broad diversity of orbital properties \citep{Udry2007,Winn2014}, expected to be fossil traces of the formation process of these systems and potentially linked as well to the properties and evolutionary stages of their host stars and their environments.

Models of planetary formation were at first developed based on the system we know best, the Solar System, and have evolved significantly in the past twenty years with the increasing flow of information derived from observations of exoplanet systems. The observed diversity of planet properties finds its origin in the physical process at play coupled with the local conditions during the formation of the system. Today, two main competing paradigms are proposed for planet formation: the core accretion model: a dust-to-planet bottom-up scenario \citep[e.g.,][]{Lissauer1993,Pollack1996,Alibert2005}, which can lead to the formation of gas giants in a few million years, and the disk's gravitational instability \citep{Boss1997, Durisen2006}, which can form massive planets on a very short timescale of a few thousand years \citep[see][for reviews on those processes]{Helled2014,Raymond2014}. Both agree on the formation of substellar companions from the circum-stellar accretion disk, but then differ depending on the initial environmental conditions in the disk, planet-disk, and planet-planet interactions.

\begin{table*}[t]
\centering
\begin{threeparttable}
\caption{Planet-search programs monitoring evolved stars.}
\begin{tabular}{|l|l|}
\hline 
\hline 
Survey & References \\ 
\hline 
the Lick G- and K-giant survey & \citet{Frink2001, Hekker2006b} \\ 
the ESO planet search program & \citet{Setiawan2003a} \\ 
the Okayama Planet Search Program & \citet{Sato2005} \\ 
\hspace{15pt}with the collaborative survey "EAPS-Net" & \citet{Izumiura2005} \\ 
the Tautenburg observatory Planet Search & \citet{Hatzes2005, Dollinger2007a} \\ 
Retired A-stars and their companions & \citet{Johnson2006} \\ 
the CORALIE \& HARPS search in open clusters & \citet{Lovis2007} \\ 
\hspace{15pt}with the follow-up program & \citet{DelgadoMena2018} \\
the Penn States Torún Planet Search & \citet{Niedzielski2007} \\ 
\hspace{15pt}with the follow-up program Tracking Advanced Planetary Systems & \citet{Niedzielski2015b} \\ 
the BOAO K-giant survey & \citet{Han2010} \\ 
the Pan-Pacific Planet Search & \citet{Wittenmyer2011}  \\ 
the Exoplanet aRound Evolved StarS project & \citet{Jones2011} \\ 
the Boyunsen Planet Search & \citet{Lee2011} \\ 
\hline 
    \end{tabular}
\label{tab:surveys}
\end{threeparttable}
\end{table*}

% Interest for companions of massive stars
Large planet-search surveys first focused on solar-type and very low-mass stars, leaving aside more massive stellar hosts that were more complex to observe and study. As the impact of the stellar mass on planet formation is still debated, it is of great interest to study the population of planets around intermediate-mass stars, that is in the $1.5-5\,M_{\odot}$ range.
Such systems are especially useful to probe the two main competing formation models. In the early phase of the process, the stellar mass seems to have little effect on the protoplanetary disk formation and evolution; however, after 3 million years stars with masses $> 2\,M_{\odot}$ start showing significant differences compared to lower mass stars, such as stronger radiation fields and higher accretion rates \citep{Ribas2015}. These impact the evolution of protoplanetary disks significantly, and by $\sim$10\,Myr there are no more disks around those higher mass stars. The typical timescale of core accretion could become problematic for massive stars that have shorter disk lifetimes \citep{Lagrange2000}. However, searching for planets orbiting main-sequence stars of intermediate masses (A to mid-F types) proves to be a challenge for Doppler searches, mainly due to the too few absorption lines present in early-type dwarfs as a consequence of their high effective temperatures, and secondly due to the rotational broadening of the lines (typical rotational velocities of 50-200\,km\,s$^{-1}$ for A-type stars, 10-100\,km\,s$^{-1}$ for early-F stars \citep{Galland2005}). A method to extract the radial-velocity from the spectrum in Fourier space was developed by \citet{Chelli2000} and then adapted and applied to early-type stars by \citet{Galland2005}. The typical radial-velocity uncertainties obtained were on the order of 100-300 and 10-50\,m\,s$^{-1}$ (normalized to a signal-to-noise ratio (S/N) = 200) for A- and F-type stars, respectively. With this technique, the team confirmed the existence of the known planet around the F7V star HD\,120136 (Tau Boo) announced by \citet{Butler1997}. The orbital parameters from \citet{Galland2005} were consistent with the values previously found. These results confirmed the accuracy of the computed radial-velocities and the possibility to detect companions in the massive, giant planetary domain for A- and F-type stars with substantial v\,sini, using high-resolution, stable spectrographs such as HARPS.

On the other hand, stars will inflate during their evolution toward the red giant branch (RGB). The effective temperature thus reduces significantly, making many more absorption lines visible in the spectra. Moreover, the rotation of the star slows down, reducing the broadening effect on the lines and making them sharper. Those stars are thus suitable bright proxies for radial-velocity planet searches around intermediate-mass stars. The analysis and interpretation of the variability of radial-velocity time series of giant stars can, however, be challenging because of their intrinsic variability, which, moreover, can also be periodic \citep[e.g.,][]{Hekker2006a,Hekker2007b}. Disentangling stellar from potential planetary contributions represents a challenge in the search for long-period and low-mass companions around evolved stars.

% Importance of stellar parameters
Bringing observational constraints on the formation and evolution of planetary systems relies principally on the determination of orbital parameters such as semi-major axes and eccentricities. It also relies on the knowledge of host star properties such as mass, radius, age, metallicity, and the abundances of individual elements. For giant stars, the mass and age are poorly constrained with the classical method of isochrone fitting because the evolutionary tracks in the HR diagram are too close to each other. The uncertainties on the observations (stellar magnitudes and colors) and systematics of the models lead to typical relative uncertainties on the stellar masses of 80-100\%. \citet{Lovis2007}, followed by \citet{Sato2007a} and \citet{Pasquini2012}, overcame this difficulty. They studied giant star populations in open clusters, for which better ages can be determined. This lead them to better mass estimates from stellar evolution models (\citet{Lovis2007} using the Padova models at solar metallicity \citep{Girardi2000}). More recently, GAIA DR2 unprecedented homogeneous photometric and astrometric data covering the whole sky allows for more precise age determination \citep[e.g.,][derived parameters such as age, distance modulus, and extinction for a sample of 269 open clusters]{Bossini2019}.

% Begining of giants survey (Cochran et ...)
In the late 1980s, a few surveys monitored giant star activity to better understand the origin of the observed large radial-velocity variations for late giant stars, with periods from days \citep{Hatzes1994} to hundreds of days \citep{Walker1989,Hatzes1993,Hatzes1999} and amplitudes of hundreds of m\,s$^{-1}$ \citep{Udry1999}. Such a variability can be explained by a combination of stellar intrinsic activity such as oscillations, pulsations or the presence of substellar companions. Several surveys searching for stable reference stars reported that many giants show relatively small radial-velocity standard deviations for early giant stars:  $\sigma_{RV} \leq$ 20\,m\,s$^{-1}$ for several among the 86 K giants followed by \citet{Frink2001} and the 34 K giants observed by \citet{Hekker2006b}. These results show that giant stars are suitable for the detection of substellar companions with radial-velocity measurements. It is worth mentioning the case of Gamma Cep here \citep{Campbell1988}, which is also considered as intrinsically variable despite the K1 giant spectral type. 

% Giants radial-velocity surveys
\citet{Frink2002}, as part of a radial-velocity survey of K giants \citep{Frink2001}, announced the detection of a 8.9\,M$_{jup}$ (minimal mass) companion orbiting the K2 III giant $\iota$ Draconis with a period of 536\,d period and an eccentricity of 0.70, making it the first substellar companion discovered orbiting a giant star. Since then, several radial-velocity surveys have been launched to follow evolved stars with intermediate masses. The list of these programs is presented in \autoref{tab:surveys}. As of November 2020, they have led to the discovery of 186 substellar companions around evolved stars in 164 systems\footnote{The list was established from the NASA Exoplanet Archive, accessible at \href{https://exoplanetarchive.ipac.caltech.edu/cgi-bin/TblView/nph-tblView?app=ExoTbls\&config=PSCompPars}{https://exoplanetarchive.ipac.caltech.edu}, by selecting the hosts with log\,g\,$\leq$\,3.5\,cm\,s$^{-2}$. The list thus contains giant and subgiant hosts.}.

\begin{figure}[t!]
        \centering
        \includegraphics[width=1\columnwidth]{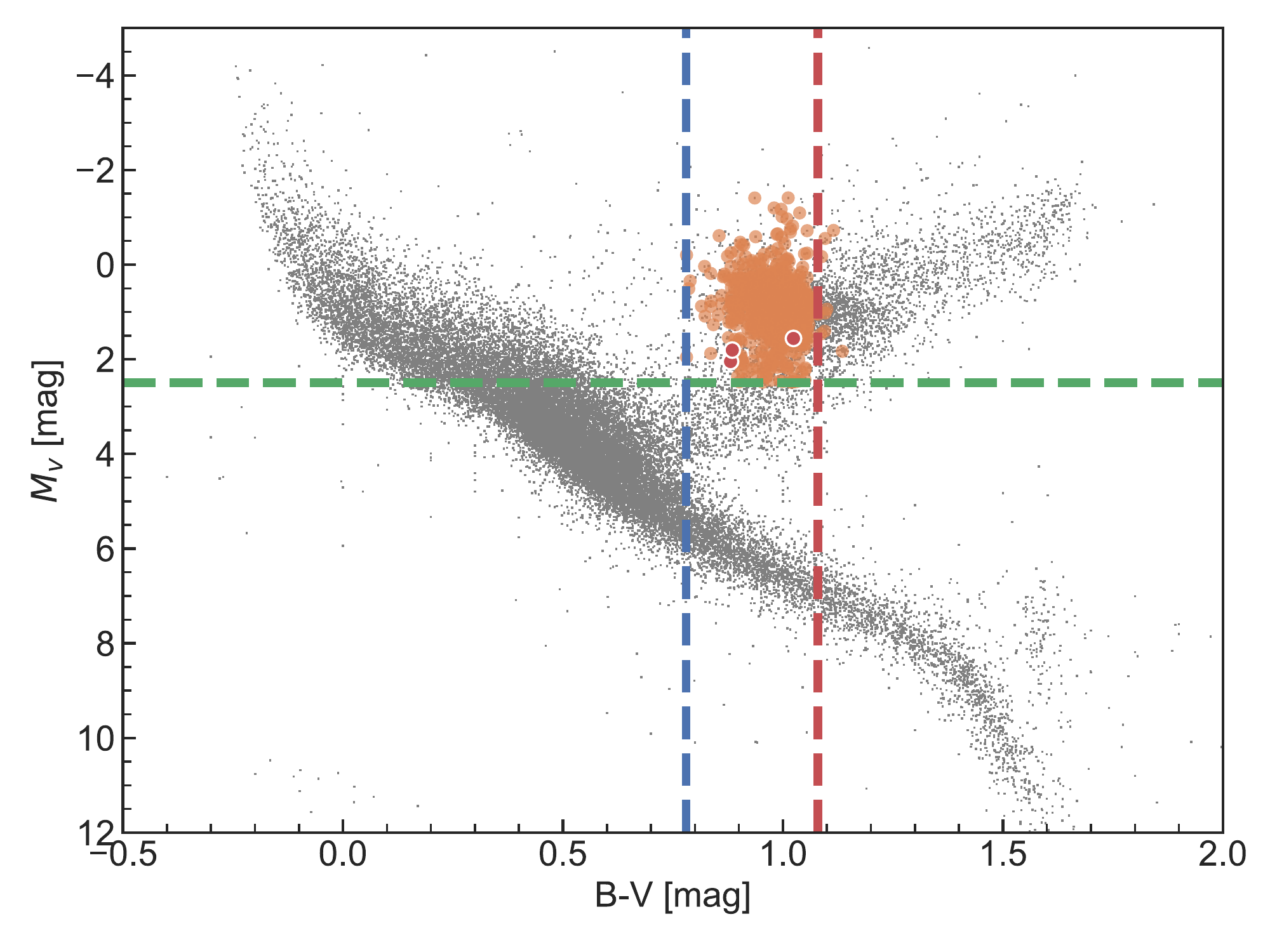}
        \caption{Color-magnitude diagram from Hipparcos measurements \citep{ESA1997} for stars with parallax precisions better than 14\% (in gray), with clear localization of our sample (in orange). The dashed lines represent the selection limits at absolute magnitude $M_V < 2.5$ (green), and between the lower (red) and higher $B-V$ values at 0.78 and 1.06. The planet host candidates presented in this paper (HD\,22532, HD\,64121, and HD\,69123) are highlighted in red.}
        \label{fig:hr_hipp}
\end{figure}

% Goal and paper structure
In this context, a survey of a volume-limited sample of evolved stars of intermediate masses, the CORALIE radial-velocity Search for Companions ArounD Evolved Stars (CASCADES), was initiated in 2006. The sample is presented in \autoref{sec:survey}, with the methods used to derive the stellar parameters described in \autoref{sec:methods}. \Autoref{sec:obs_analysis} describes the complete time series acquisition process and analysis, from the search for periodicities in the activity indicators to the Keplerian fitting of the radial velocities, which lead to orbital solutions of three newly discovered planetary companions in \autoref{sec:results}. Additional detections and statistical analysis of the survey will be presented in a series of subsequent publications. Finally, in \autoref{sec:conclusion} we discuss some implications of the first discoveries and provide concluding remarks in the broader context of the population of giant stars hosting substellar companions.

%--------------------------------------------------------------------
\section{The CASCADES survey}\label{sec:survey}

\subsection{Goals and sample definition}\label{sub:sample_def}

\begin{figure}[th!]
        \centering
        \includegraphics[width=1\columnwidth]{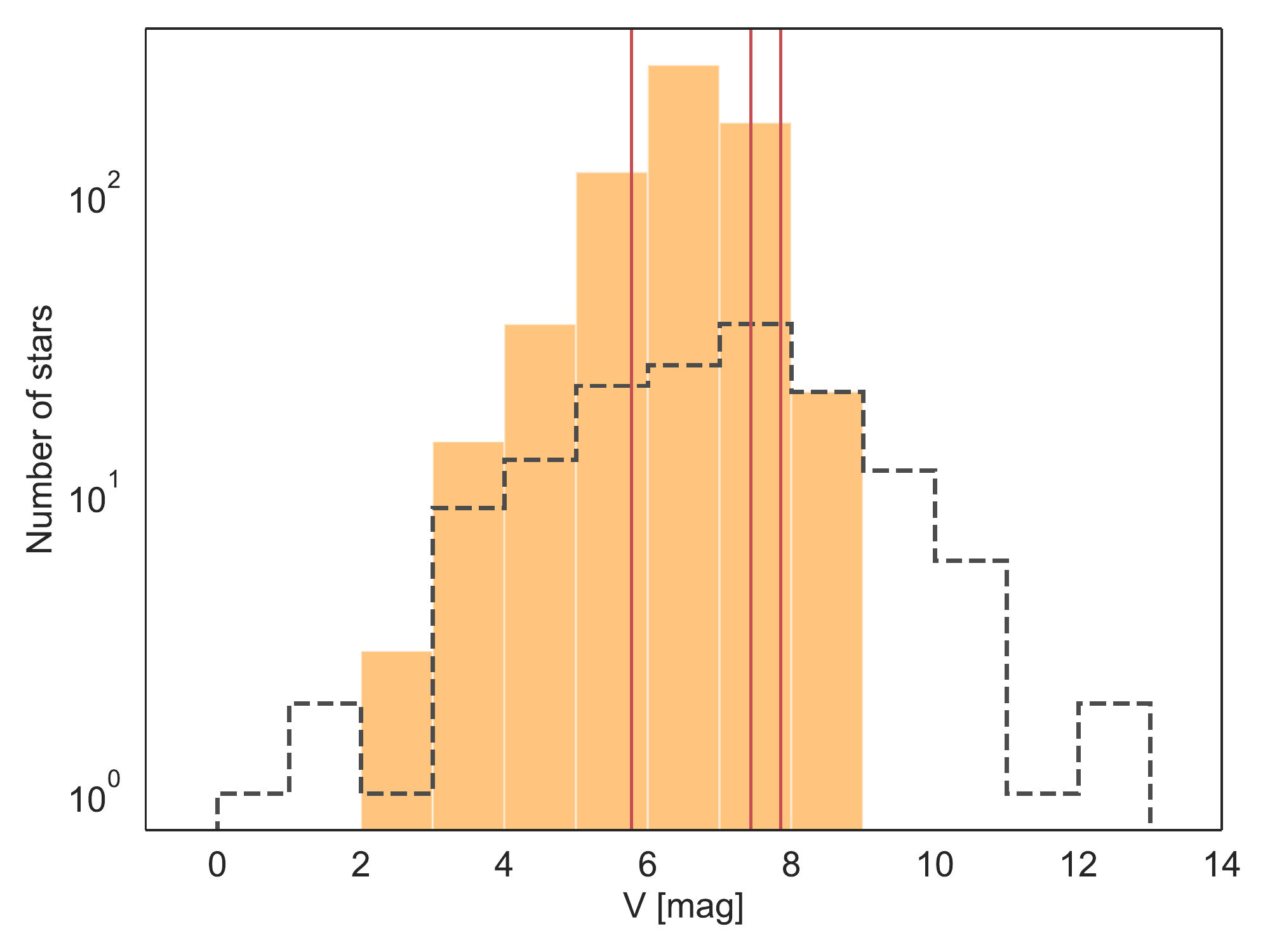}\\
        \includegraphics[width=1\columnwidth]{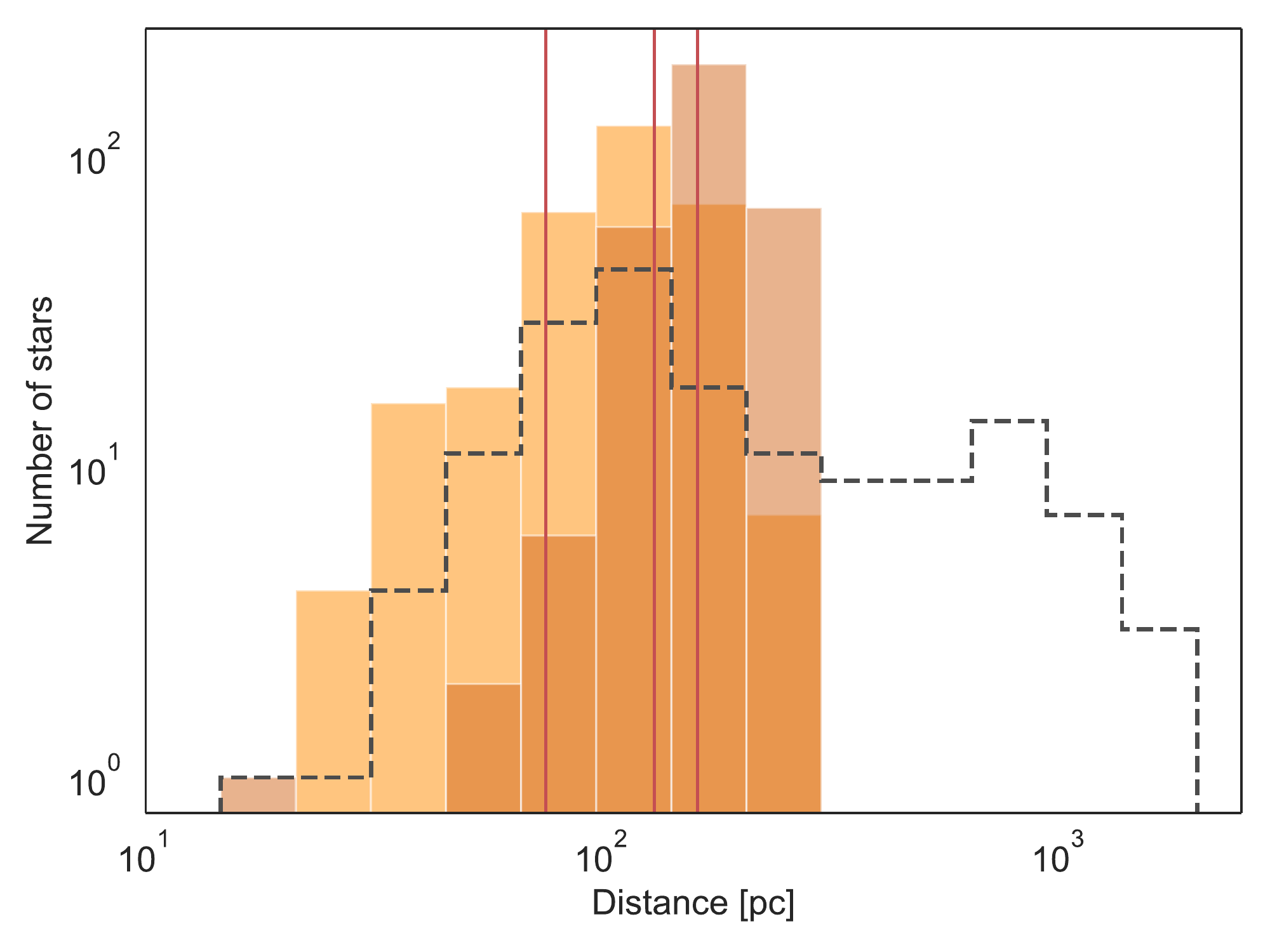}
        \caption{Distributions of CASCADES survey sample (filled orange histogram), compared with the published giant stars known to host planet companions (dashed histogram). The top panel displays the apparent magnitude (TYCHO-2 catalog, \citet{Hog2000}) and the bottom one the distance (GAIA DR2, \citet{Bailer-Jones2018}). The CASCADES original 2006 sample and its 2011 extension are differentiated by lighter and darker orange shades, respectively. The positions of HD\,22532, HD\,64121, and HD\,69123 are represented by red lines.}
\label{fig:hist_v}
\end{figure}

In the context described above, in 2006 we (i.e., Christophe Lovis and Michel Mayor) launched a precise radial-velocity survey of evolved stars of intermediate masses, which we refer to as "giant stars" in this paper. The main motivation was to better understand the formation of planetary systems and their evolution around stars more massive than the Sun by completing the existing statistical properties of giant host stars and their companions. To conduct a well-defined statistical study, we chose the following criteria for the definition of the sample:

\begin{figure}[th!]
        \centering
        \adjincludegraphics[width=1\columnwidth, trim={0 0 0  {-.042\height}},clip]{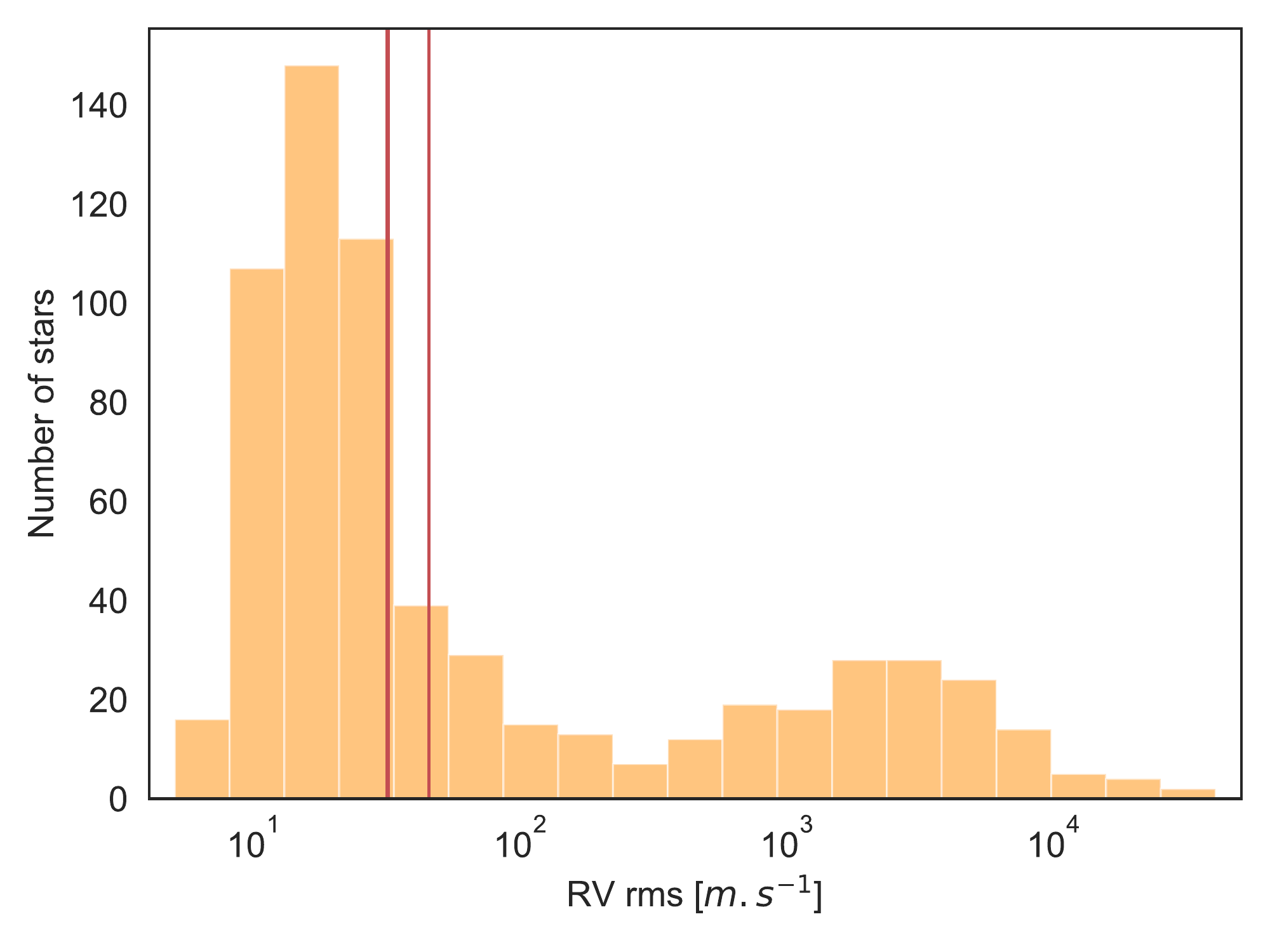}\\
        \includegraphics[width=1\columnwidth]{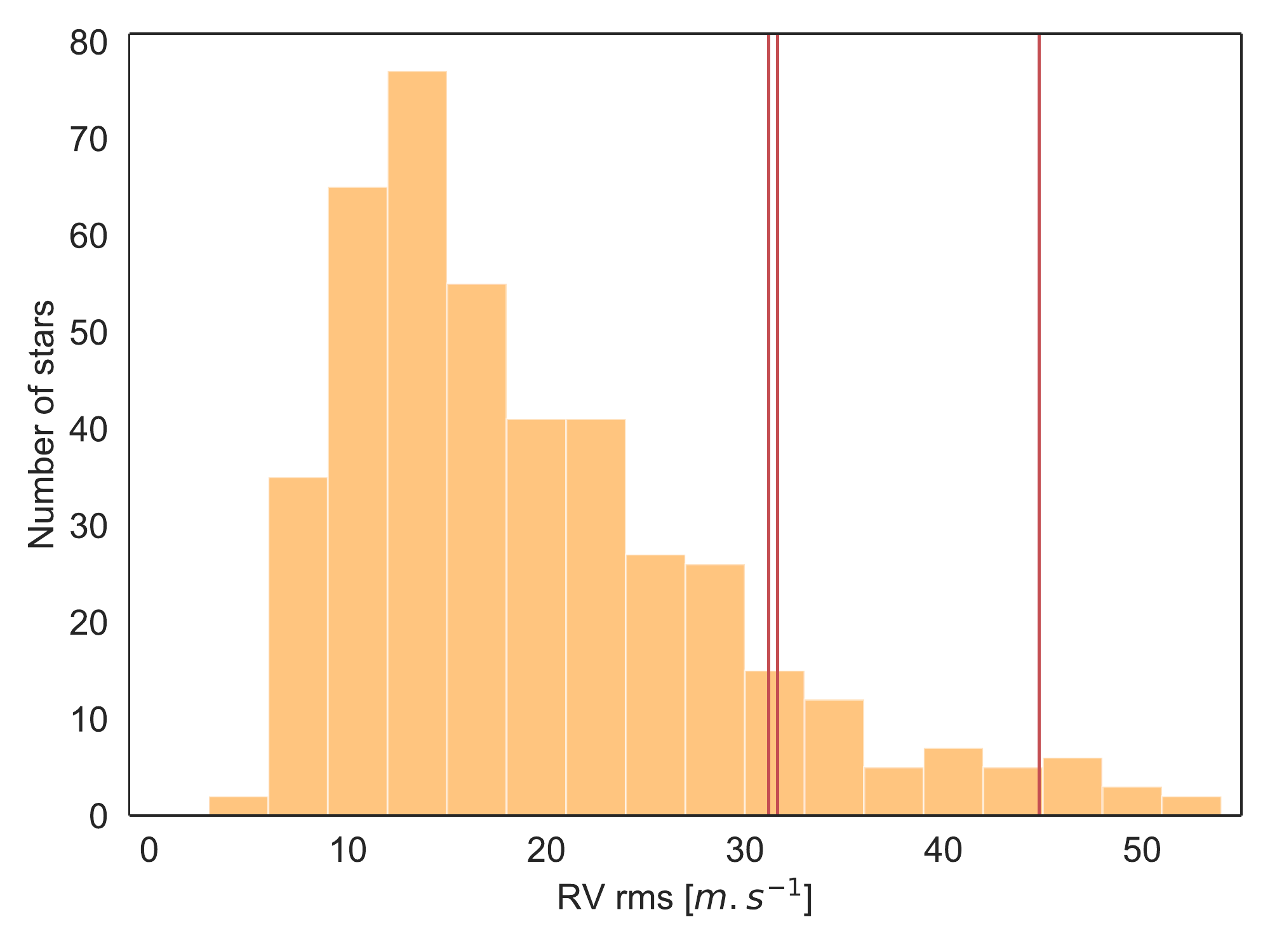}
        \caption{Distribution of the radial-velocity dispersion observed for the stars in the sample. Top: Full range in logarithmic scale. Bottom: Zoomed-in image of smaller values, in linear scale. The positions of HD\,22532, HD\,64121, and HD\,69123 are represented by red lines.}
\label{fig:hist_rms}
\end{figure}

We defined a volume-limited sample, $d\leq300$\,pc. The selection was done in 2005 from the Hipparcos catalogue \citep{ESA1997}. We only selected stars from the catalog with a precision on the parallax better than 10\,\%. To increase the statistics, the sample was extended in 2011 to targets with a parallax precision better than $14\,\%$. We selected stars with $M_V < 2.5$ and $B-V>0.78$, to avoid stars on the main sequence. We selected only early-type giants, with G and K spectral types and luminosity class III. To avoid later types, which are known to be intrinsically variable, we introduced a color cut-off at $B-V<1.06$. We avoided close visual binaries for which we might have contamination by the secondary in the spectrograph fiber. The limit on the separation was set at $6''$. Finally, we selected stars observable by Euler, in the southern hemisphere, with a declination below $-25^{\circ}$, to be complementary to existing surveys in the north reaching down to $\delta=-25^{\circ}$.

The criteria and the final sample of 641 stars are represented in \autoref{fig:hr_hipp} displaying the Hipparcos color-magnitude diagram with the selected sample highlighted. In \autoref{fig:hist_v}, we show the distribution of stellar apparent magnitudes and distances for the sample. We paid particular attention to the potential bias induced by the criterion on the parallax precision when carrying out the statistical study of the sample.
Because of its size, timespan of observations, and the quantity and quality of the measurements, the above-defined planet search program is expected to significantly improve our knowledge of planetary systems around giant stars.

\subsection{Instrument description and observations}
\label{sub:instru}

Observations for this survey began at the end of 2006 and have been conducted since then with the CORALIE spectrograph on the 1.2-m Leonard Euler Swiss telescope located at La Silla Observatory (Chile). CORALIE is a 2-fiber-fed echelle spectrograph (2" fibers on the sky). It covers the 3880-6810 Å wavelength interval with 68 orders. The spectral resolution of the instrument was originally of 50\,000. The observations are performed in the so-called simultaneous thorium mode\footnote{A thorium-argon lamp illuminates fiber B at the same time as the star is observed on fiber A. Both spectra are recorded on the CCD.}. In 2007 and 2014, CORALIE went through two significant hardware upgrades to improve the overall performance of the instrument, such as improving the throughput of the instrument (gain of 2 magnitudes) and increasing the resolution to 60\,000. A Fabry-Perrot etalon was installed as well to replace the thorium-argon lamp used to track the variation of the spectrograph during the night. For simplicity, we refer to each dataset as {\footnotesize CORALIE-98} (or COR98) for the first version of the instrument, {\footnotesize CORALIE-07} (or COR07) for the 2007 upgrade and {\footnotesize CORALIE-14} (or COR14) for the 2014 instrument upgrade. The instrumental precision evolved through these upgrades, from 5\,m\,s$^{-1}$ for COR98, to 8\,m\,s$^{-1}$ for COR07, and finally to 2.5-3.2\,m\,s$^{-1}$ for COR14. Complete information on instrumental aspects and the precisions reached are given in, for instance, \citet{Queloz2000,Segransan2010,Segransan2021}.

\begin{figure}[t!]
        \centering
        \includegraphics[width=1\columnwidth]{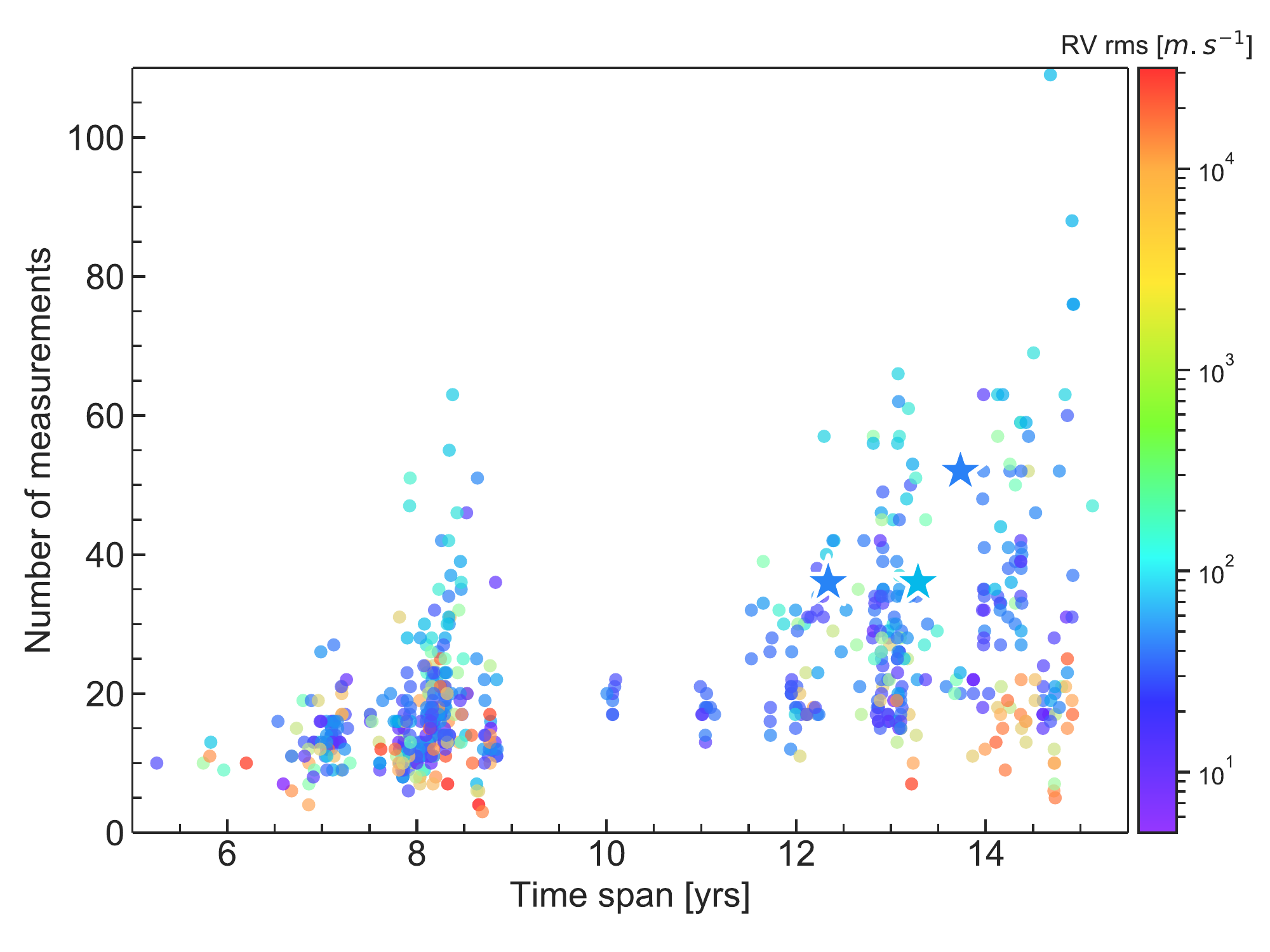}\\
        \caption{Number of measurements per star in the sample plotted against the respective time span of the observations. The points are color-coded by the radial-velocity rms. The star markers represent HD\,22532, HD\,64121, and HD\,69123.}
\label{fig:pts_tspan}
\end{figure}

Considering the size of the sample and the first results on the intrinsic variability of giant stars obtained by earlier surveys, we tried to optimize the exposure time in our program to match the limit set by stellar variation. Evolved stars often show distributions of the dispersion of radial-velocity time series ranging from $\sim$ 5\,m\,s$^{-1}$ up to a few hundreds of m\,s$^{-1}$, with a mode at $\sim$ 15\,m\,s$^{-1}$ \citep{Lovis2004,Hekker2006b,Quirrenbach2011}. One of our objectives is also to test if this intrinsic limit can be lowered by modeling stellar variability, and thus to be able to recognise low-amplitude planet signatures. To reach a sufficient precision ($\leq$ 2-3\,ms$^{-1}$), exposure times between 3 and 5 minutes are adequate for these very bright stars. 

The CASCADES survey is also interesting in the broader context of projects running on the telescope. The brightness of the stars in the sample makes them ideal back-up targets when weather conditions are unfavorable. 

Almost 15\,000 radial-velocity measurements have been obtained so far for the 641 stars of the sample. The study of the radial-velocity dispersion (see \autoref{fig:hist_rms}) confirms a distribution with a peak around 13\,m\,s$^{-1}$, with values as low as $\sim$ 4\,m\,s$^{-1}$. We also see a significant tail at higher values with a secondary peak around 20\,km\,s$^{-1}$ corresponding to binary systems. 

Our observational effort is illustrated in \autoref{fig:pts_tspan}, with the three host stars presented in this paper (located with a $\star$ symbol) being among the most observed in the sample in terms of duration and number of data points. The program is continuing to obtain a minimum of 20 points per star and so can perform a solid statistical analysis of the sample.

%--------------------------------------------------------------------
\section{Determination of stellar properties}\label{sec:methods}

\subsection{Spectrocopic parameters of the stars in the sample}\label{sub:spec_param}
The analysis of high-resolution spectra can provide reliable stellar parameters such as effective temperature $T_{eff}$, surface gravity $log\,g,$ and iron metallicity ratio $[Fe/H]$ (which we refer to as metallicity in this paper for simplicity). \citet{Alves2015} presented a catalog of precise stellar atmospheric parameters and iron abundances for a sample of 257 G and K field evolved stars, all of them part of our sample, using UVES and CORALIE spectra. The approach, based on \citet{Santos2004}, uses the classic curve-of-growth method. The equivalent width of a set of Fe I and II lines is measured and their abundances are calculated. Then, the stellar parameters are derived when excitation and ionization balances are satisfied simultaneously under the assumption of local thermodynamic equilibrium. For a detailed description of the method and the results, we direct the reader to \citet{Santos2004,Sousa2014,Alves2015}. 

\begin{figure*}[t!]
        \centering
        \includegraphics[width=.33\textwidth]{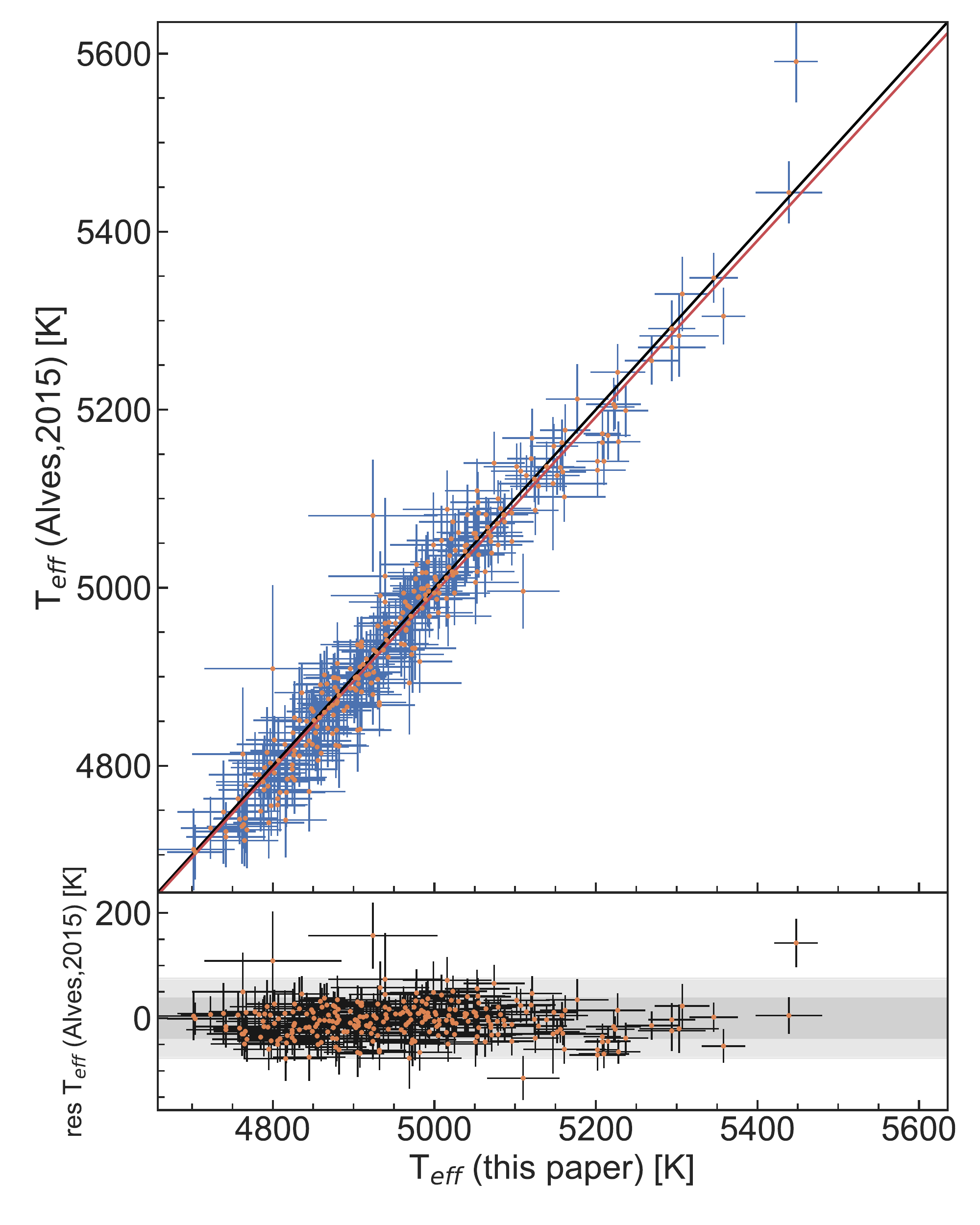}
        \includegraphics[width=.33\textwidth]{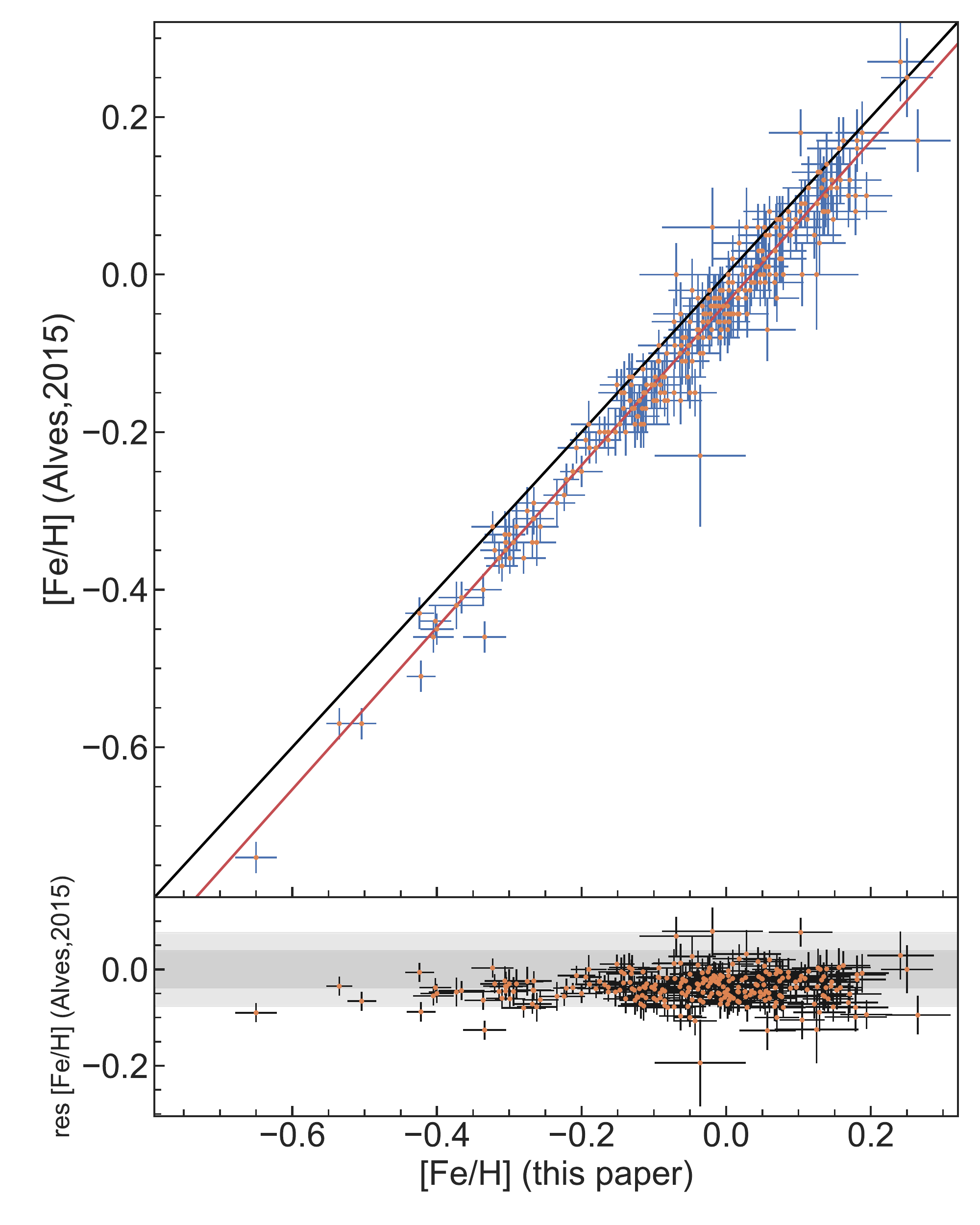}
        \includegraphics[width=.33\textwidth]{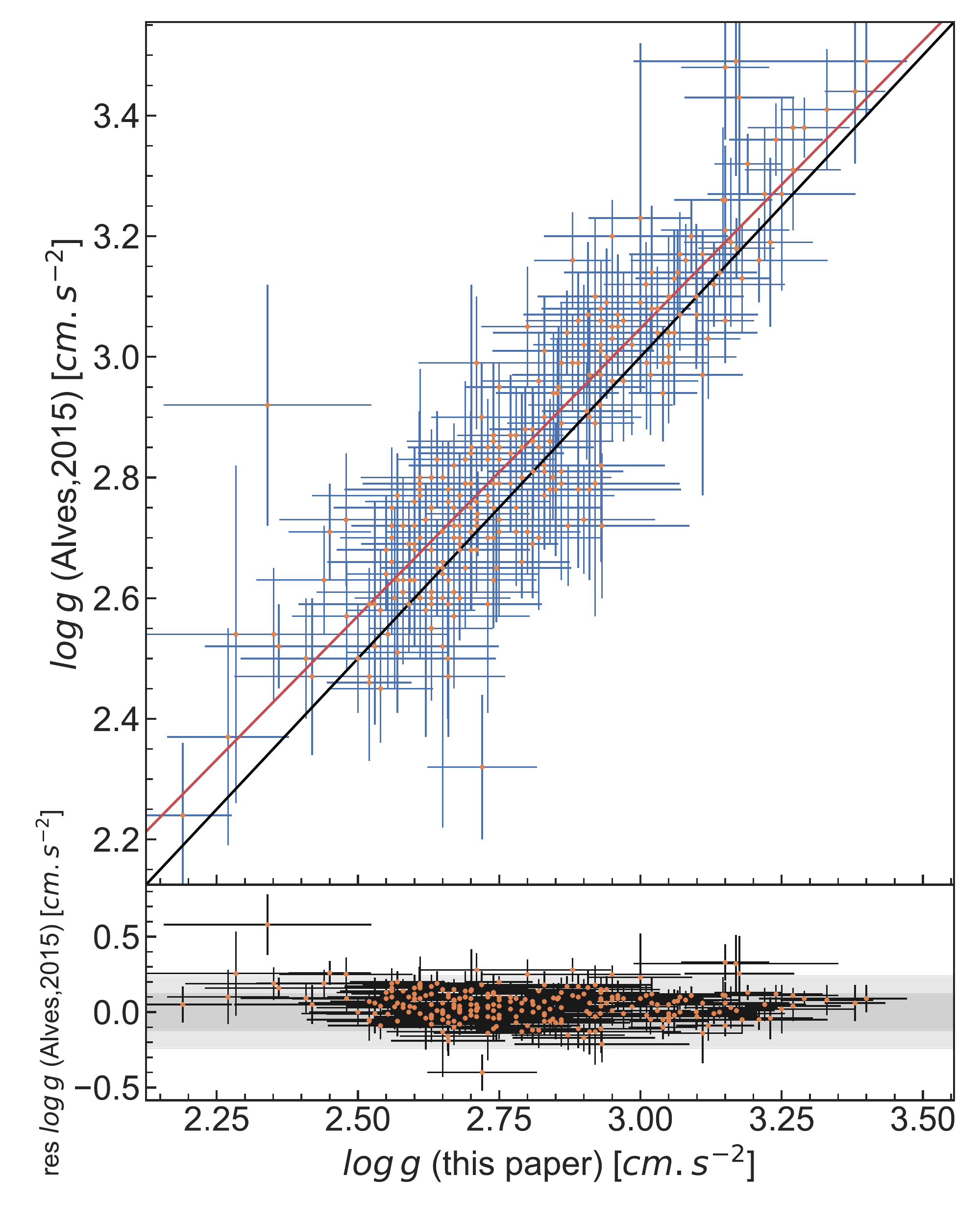}
        \caption{Comparison plots of spectroscopic parameters extracted from CORALIE (this paper) and UVES \citep{Alves2015} spectra of the subsample of 254 stars in common. The black diagonal line represents the 1:1 correlation, and the red line represents the linear fit of the data. At the bottom of each figure, the residuals compared to the 1:1 correlation are shown, with their 1 and 2\,$\sigma$ dispersions represented by the shaded regions. Left: Effective temperatures seem to be perfectly correlated, with a dispersion of $\sim38$\,K. Middle: Metallicity ratio of iron [Fe/H] shows an apparent positive offset of the order of the dispersion of the data around the 1:1 correlation, $\sim0.04$\,[dex] in favor of our estimation. More than 50\% of the subsample is inside the $1\sigma$ region and 90\% inside the $2\sigma$. Right: Logarithm of the surface gravity shows a good correlation, with an offset $\sim0.05\,cm.s^{-2}$ lower than apparent dispersion of the data around the 1:1 correlation. More than 70\% of the subsample is inside the $1\sigma$ region.}
        \label{fig:comp_param_stell}
\end{figure*}

Before 2014, the CORALIE spectra obtained for precise RV measurements were polluted by the strongest lines of the Thorium-Argon spectrum from the calibration fiber, and thus they were not suitable for a precise spectroscopic analysis. Dedicated additional spectroscopic observations (without calibration) were then obtained for our sample stars. After the CORALIE upgrade in 2014, the calibration spectrum from the Fabry-Perrot etalon was no longer polluting the stellar spectrum, and the observations of radial-velocity measurements can also be used for the spectroscopic analysis. We stacked the CORALIE-14 spectra from all stars in our sample to reach a high enough S/N (the median S/N of the master spectra is about 170), and we derived the spectroscopic parameters $T_{eff}$, $log\,g$ and $[Fe/H]$ using the ARES\citep{Sousa2007,Sousa2015} $+$ MOOG \citep{Sneden1973} methodology following \citet{Sousa2014}. For most of the stars, with T$_{eff}$\,<\,5200\,K, the iron line list presented in \citet{Tsantaki2013} was used. While for the hotter stars we used the standard line list presented in \citet{Sousa2008a}. We then compared these results with the ones presented in \citet{Alves2015} for the subsample of 254 common stars. \Autoref{fig:comp_param_stell} shows the good agreement found between the parameters obtained from the UVES and CORALIE spectra: in the case of metallicity, we observe an apparent positive offset of the order of the dispersion of the data around the 1:1 correlation, $\sim0.04$\,[dex] in favor of our estimation. More than 50\% of the subsample stars are inside the $1\sigma$ region and 90\% are inside $2\sigma$. These results confirm the quality of the CORALIE-14 spectra to extract reliable atmospheric parameters from them.

\begin{figure}[t!]
        \centering
        \adjincludegraphics[width=.49\columnwidth, trim={0 0 {.02\height} {-0.035\height}},clip]{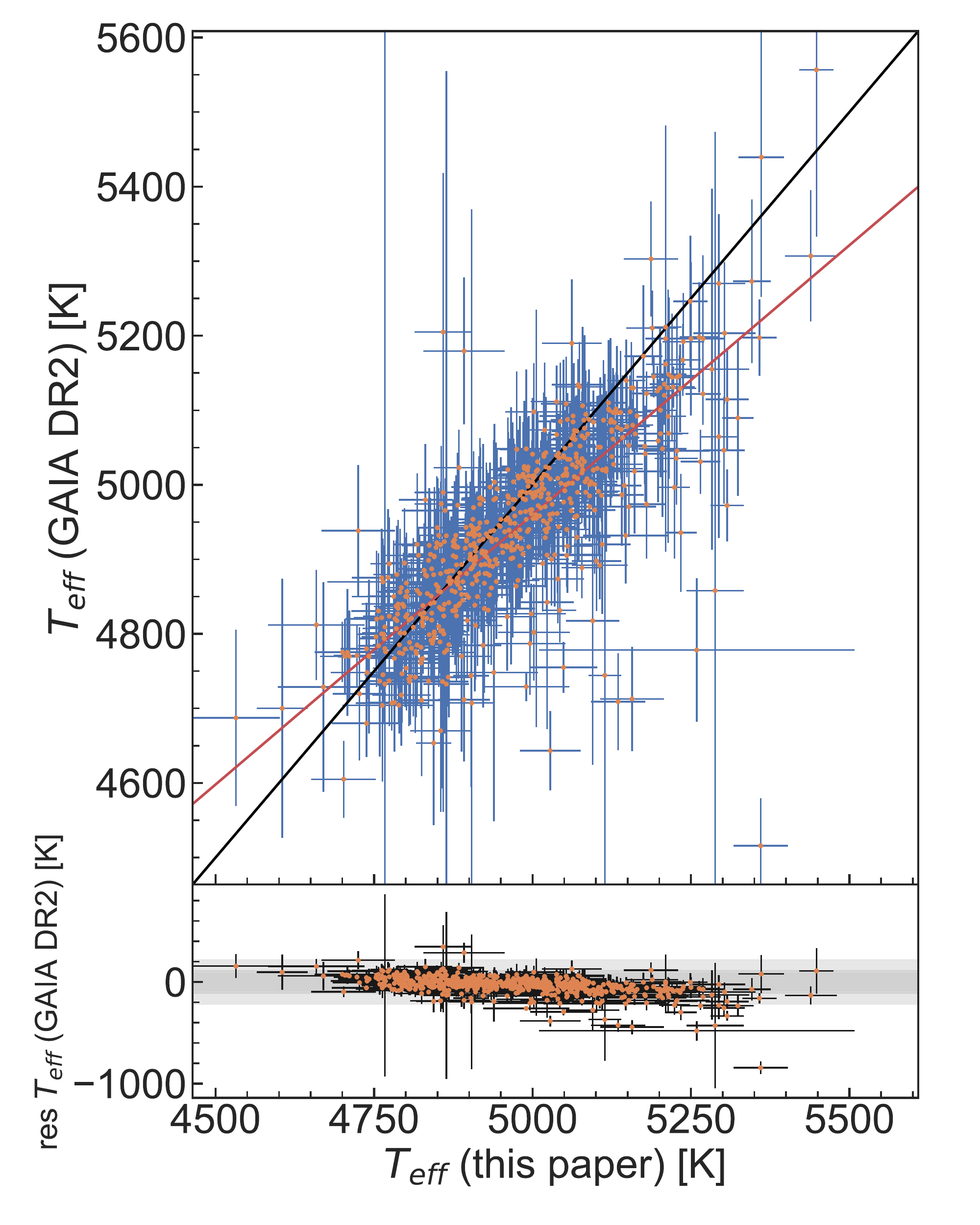}
        \adjincludegraphics[width=.49\columnwidth, trim={{.02\height} 0 0 {-0.035\height}},clip]{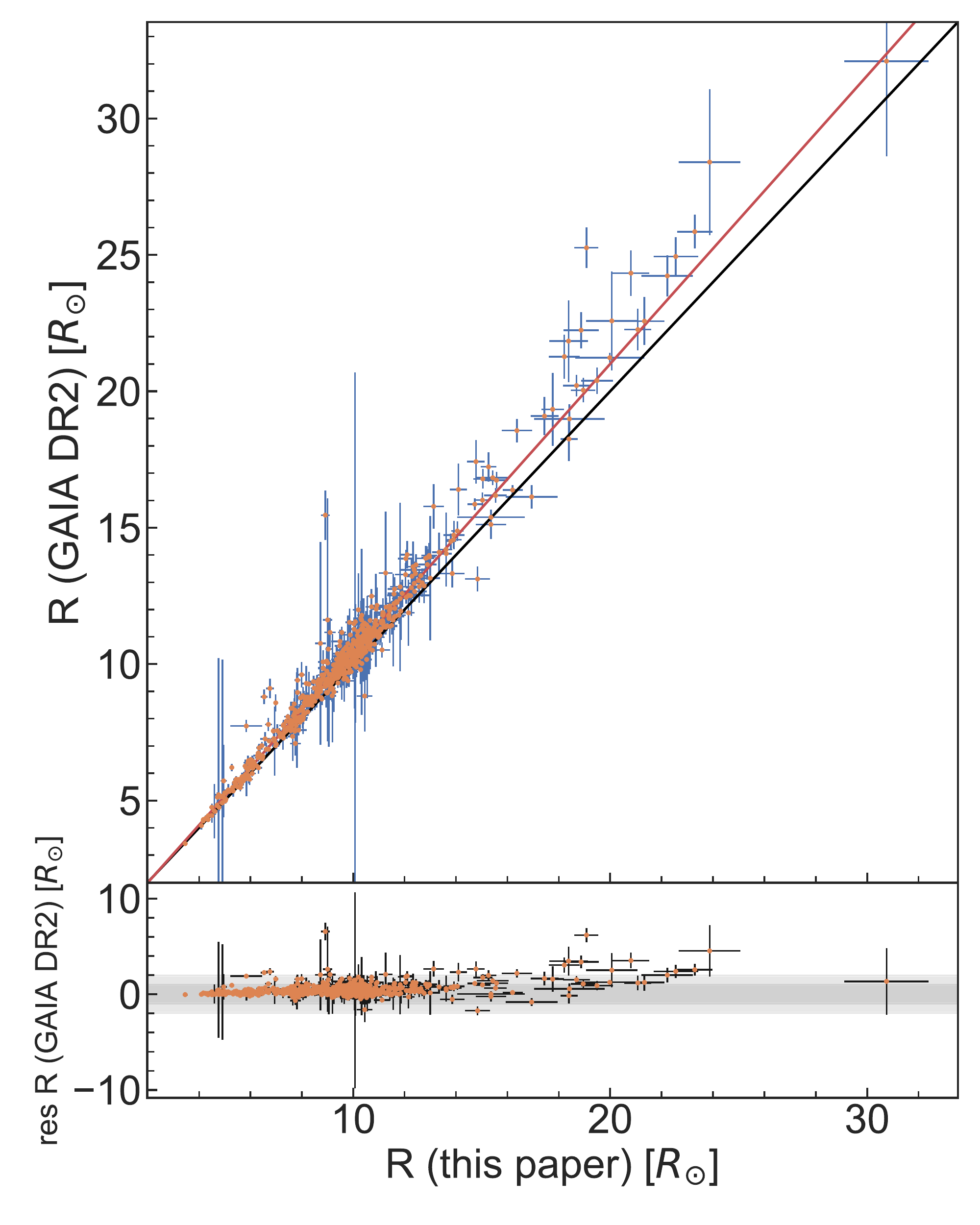}
        \caption{Comparison plots of photometric (from \citep{Brown2018}) and spectroscopic (from this paper) determinations of the effective temperatures and the stellar radii. The black diagonal line represents the 1:1 correlation and the red line represents the linear fit of the data. At the bottom of each figure, the residuals compared to the 1:1 correlation are shown, with their 1 and 2\,$\sigma$ dispersions represented by the shaded regions. Left: Effective temperatures seem to show a linear trend, but this is not significant compared to the dispersion of the date of $\sim110$\,K, inside which more than 85\% of the data is located. Right: Radii are in good agreement but show an apparent trend and increasing dispersion for masses above than $\sim15\,M_{\odot}$.}
        \label{fig:comp_spec_gaia}
\end{figure}

\begin{figure}[t!]
        \centering
        \includegraphics[width=1\columnwidth]{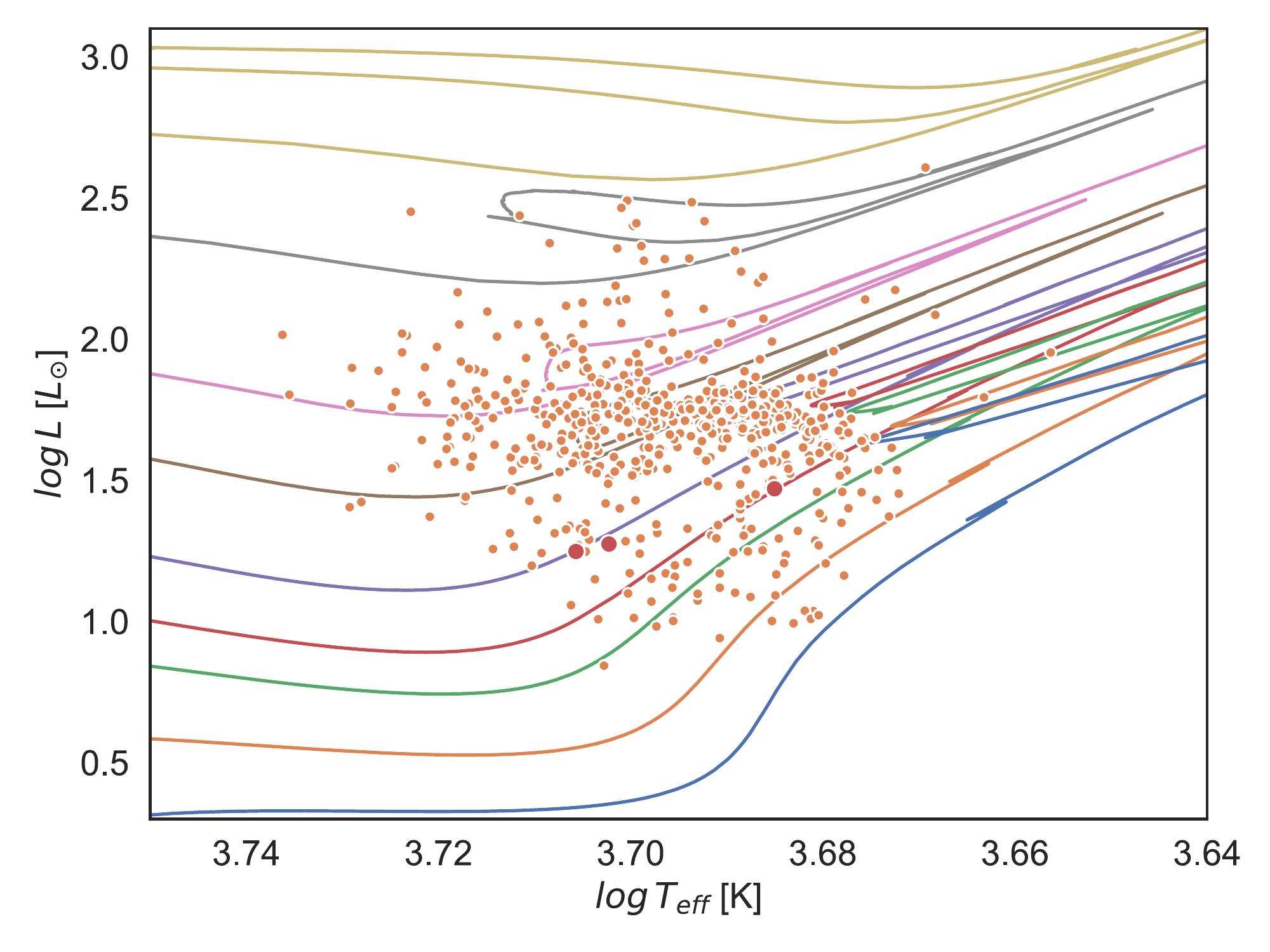}
        \caption{Positions in Hertzsprung-Russell diagram of the subsample of 620 stars for which we derived spectroscopic parameters. The three host stars we focus on in the present paper are highlighted as red dots. We adopted the luminosities obtained with the method described in \autoref{sub:stellar_mass}. The evolutionary tracks are from models of \citet{Pietrinferni2004} for different stellar masses ($1.0$, $1.2$, $1.5$, $1.7$, $2.0$, $2.5$, $3.0$, $4.0,$ and $5.0$ $M_{\odot}$ from bottom to top). They are for models with solar metallicity.}
        \label{fig:evoltrack}
\end{figure}

\subsection{Stellar luminosities, radii, and masses}\label{sub:stellar_mass}
% say something about the bayesian method developeed by Bailer-jones (2018). see the sentence in Katz et al 2018 publication (https://ui.adsabs.harvard.edu/abs/2018A%26A...616A..11G/abstract) : "We established that inverting the parallax leads to unbiased distances out to about 1.5 kpc, with overestimates of the order of 17\% at 3 kpc. We therefore have to bear in mind that the distance bias in the extremes of our main sample is non-negligible." and "Alternatively, Bayesian methods might be used to infer distances from parallaxes instead of selecting stars with small relative uncertainty" 
We derived the luminosity $L$ for the stars in our sample using the Gaia DR2 parallaxes corrected by \citet{Bailer-Jones2018}\footnote{\citet{Bailer-Jones2018} provided purely geometric distance estimates by using an inference procedure that accounts for the nonlinearity of the transformation (inversion of the parallax) and the asymmetry of the resulting probability distribution.}, $V$-band magnitudes from \citet{Hog2000}, and the bolometric correction relation $BC$ of \citet{Alonso1999}\footnote{Considering the short distance of the stars of the sample, the extinction was not taken into account.}. We then used this luminosity and the spectroscopic effective temperature $T_{eff}$ to compute the stellar radii using the Stephan-Boltzmann relation. The uncertainties of the radii were estimated using a Monte Carlo approach. We compared our temperatures and radii with the GAIA DR2 values and found them to be in good agreement, as illustrated in \autoref{fig:comp_spec_gaia}.

% \subsubsection{Distribution of the masses of the sample}

The derived luminosities and spectroscopic effective temperatures are plotted in the Hertzsprung-Russell diagram in \autoref{fig:evoltrack}, together with the stellar evolutionary tracks at solar metallicity of \citet{Pietrinferni2004}\footnote{Available on the BaSTI database \href{http://basti.oa-teramo.inaf.it/index.html}{http://basti.oa-teramo.inaf.it.}}. Those were used to estimate the masses of our stars using the SPInS software \citep{Lebreton2020}\footnote{\href{https://dreese.pages.obspm.fr/spins/index.html}{https://dreese.pages.obspm.fr/spins/index.html}, which employs a global Markov Chain Monte Carlo (MCMC) approach taking into account the different timescales at various evolutionary stages and interpolation between the tracks.}. The approach compares the luminosity, effective temperature, logarithm of surface gravity, and [Fe/H] of individual objects to theoretical evolutionary tracks and accounts for the observational errors in these four quantities. Giant stars are located in the area of the Hertzsprung-Russell diagram where individual evolutionary tracks are close to each other; thus, the derived precisions on the stellar masses might be overestimated. However, in our sample, the degeneracy between the horizontal branch (HB) and RGB is not too pronounced. Comparisons with masses derived from detailed asteroseismic modeling (see \autoref{subsec:masses_astero}) show some small differences. These discrepancies mainly originate from the use of a fixed enrichment law in the grid of \citet{Pietrinferni2004}, such that the chemical composition a the stars of our sample is fully determined from the determination of $\left[Fe/H\right] $ in our modeling using SPInS. This was not the case for the seismic modeling pipeline, where the additional constraints justified allowing for an additional free parameter. Thus, solutions with a chemical composition deviating from a fixed enrichment law where helium and metal abundances are tied together were possible outcomes of the modeling procedure. This is particularly visible for the two stars with Fe/H $\sim-0.2$ studied in \autoref{subsec:masses_astero} (HD\,22532 and HD\,64121), for which the asteroseismic analysis reveals an initial helium abundance slightly higher than the solar one despite their subsolar metallicity. This situation of course cannot be reproduced by models with an initial helium abundance fixed by the metallicity, which then leads to an overestimation of the stellar mass determined by SPInS to compensate for the incorrect helium abundance and reproduce the observed location in the HR diagram.

\begin{figure*}[t!]
        \centering
        \adjincludegraphics[width=1\textwidth, trim={0 0 0 0},clip]{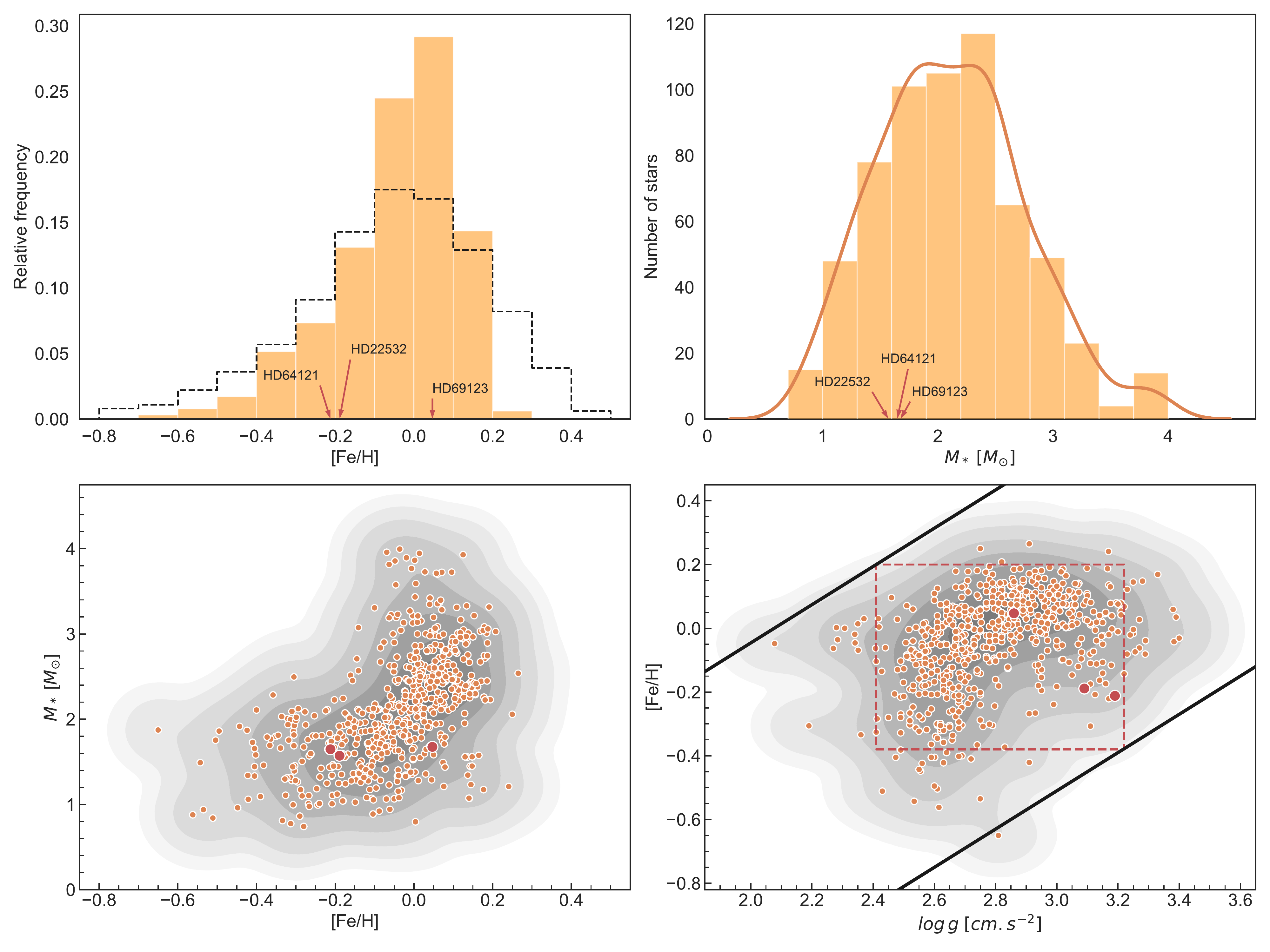}
        \caption{Distributions and relations between stellar parameters for our subsample of 620 stars. Top left: Metallicity distribution of the stars of our sample (colored in orange) compared with the same distribution for the $\sim1000$ stars (dashed histogram) in the CORALIE volume-limited sample \citep{Udry2000,Santos2001}. Top right: Distribution of the stellar masses obtained from track fitting. The corresponding kernel density estimation is overplotted in orange, using a Gaussian kernel. Bottom left: Mass vs. metallicity relation. (bottom right) Metallicity vs. logarithm of surface gravity. The two black lines were drawn by eye and show the biases in the samples due to the B-V cut-off. The red-dashed rectangle delimits the area of the potential unbiased subsample. The three planet hosts presented in this paper are represented by red dots.}
        \label{fig:params_distrib}
\end{figure*}

Nevertheless, we can consider that these fits still allow us to globally estimate the stellar masses of our sample of stars. The obtained distribution for all stars is shown in \autoref{fig:params_distrib}. The median of the distribution is found around $2.1$ $M_{\odot}$, corresponding to intermediate-mass stars. However, as mentioned above, some uncertainty on the evolutionary stage of a subsample of stars of our sample could have affected our determinations, especially if we consider deviations from a given chemical enrichment law. Indeed, some stars of our sample identified here as being around $\approx 2.0$ $M_{\odot}$ on the RGB could actually be low-mass stars in the red clump. This degeneracy could be lifted using asteroseismic observations of dipolar oscillation modes that allow us to unambiguously determine the evolutionary stage of these stars \citep{Beck2011,Bedding2011}.

\Autoref{tab:sample_table} shows example lines of the complete set of stellar parameters for our sample, available online\footnote{Available at \href{https://cdsweb.u-strasbg.fr}{CDS} and \href{https://dace.unige.ch/catalogs/?catalog=CASCADES}{https://dace.unige.ch.}}. We also illustrate our results in \autoref{fig:params_distrib}. The metallicity distribution decreases with increasing [Fe/H] for [Fe/H]\,>\,0.0-0.1, similarly to the metallicity distribution \citep{Santos2001} of a large, volume-limited sample of dwarf stars in the solar neighborhood, included in the CORALIE survey \citep{Udry2000}. We observe that our sample of giants is lacking the metal-rich and very metal-poor stars. This tendency has been observed in many studies \citep[e.g.,][]{Luck2007,Takeda2008,Ghezzi2010,Adibekyan2015a,Adibekyan2019}. It may be related to the fact that giants, most of them being more massive, are younger than their dwarf counterparts. They thus do not have time to migrate far from the inner to the outer disks of the galaxy during their short lifetimes \citep{Wang2013,Minchev2013}. \citet{Adibekyan2019} also addresses the role of the age-metallicity dispersion relation \citep{DaSilva2006,Maldonado2013}, as well as potential selection effects due to B-V color cut-off \citep{Mortier2013}, which excludes low-log-g stars with high metallicity and high-log-g stars with low metallicity. We illustrate this effect in \autoref{fig:params_distrib}, in the same way as \citep{Adibekyan2015a}, by drawing diagonal lines that show the biases in the sample due to the color cut-off. We also represent the area that would correspond to an unbiased subsample inside a cut rectangle (red dashed rectangle). We will perform a detailed statistical study of the stellar parameters of our sample in future work.

\subsection{Asteroseismic masses for the three planet hosts}\label{subsec:masses_astero}
To go further and improve the mass estimation for the host star, we performed a detailed seismic analysis of the TESS short-cadence photometric data \citep{Ricker2014}, following the methodology of \citet{Buldgen2019}. This asteroseismology approach has the considerable advantage of leading to a highly precise and accurate mass estimate independently of any stellar evolution models. We used the method to extract masses for the three stellar hosts presented in this paper (see \autoref{tab:stellar_params}) to obtain a better estimation of the minimum mass of their planetary companions. The seismic masses can also be used as a benchmark to assess the accuracy of the masses obtained from evolutionary models, which in this case appear to be overestimated by an offset of $\sim$0.3-0.4\,M$_\odot$ but are consistent within 3-4 $\sigma$. This aspect will be addressed in a forthcoming paper once more asteroseismic masses are available. In practice, the mass estimates we present here are a result of the combination of seismic inversions of the mean density with the values of the stellar radii derived from GAIA parameters. The seismic inversion of the mean density was carried out following the methodology of \citet{Buldgen2019} and validated on eclipsing binaries. This estimate still depends on the seismic data, as well as the details of the determination of the radii from GAIA and spectrocopic data, such as the bolometric corrections and extinction laws. An in-depth description of the data extraction and seismic modeling, as well as an analysis of the orbital evolution and atmospheric evaporation of the planetary systems, can be found in a companion paper \citep{Buldgen2021}.\\
In \autoref{tab:stellar_params}, we summarize the spectroscopic parameters of the three stellar hosts announced in this paper and their masses derived from evolutionary tracks and asteroseismology.

% Short Online Table stellar parameters
\begin{table*}[!ht]
\caption{Example entries of the table of stellar parameters for the complete sample, available online at CDS.$^8$}
\begin{center}
\scalebox{0.562}{%
\begin{tabular}{l*{19}{c}}        
\hline
\hline
HD & Sp. Type & $V$ & $B-V$ & $BC$ & $\pi$ & $d$ & $M_{V}$ & $Bp-Rp$ & $G$ & $T_{eff}$ & $log\,g$ & $[Fe/H]$ & $M_{*}$ & $L_{*}$ & $R_{*}$ \\
& & [mag] & [mag] & $BC$ & [mas] & [pc] & [mag] & [mag] & [mag] & [K] & [cm\,s$^{-2}$] & [dex] & [M$_{\odot}$] & [L$_{\odot}$] & [R$_{\odot}$]  \\
& [1] & [2] & [2] & [3] & [4] & [5] & [2,4,5] & [4] & [4] & [6] & [6] & [6] & [7] & [2,3,4] & [2,3,4,6] \\
\hline
496 & K0III & 3.88 $\pm$ 0.01 & 1.00 $\pm$ 0.01 & -0.312 $\pm$ 0.016 & 24.20 $\pm$ 0.29 & 41.3 $\substack{+0.5 \\ -0.5}$ & 0.80 $\pm$ 0.03 & 1.218 $\pm$ 0.006 & 0.42 $\pm$ 0.03 & 4858 $\pm$ 41 & 2.56 $\pm$ 0.10 & -0.01 $\pm$ 0.03 & 1.93 $\pm$ 0.23 & 50.44 $\pm$ 1.46 & 10.03 $\pm$ 0.22 \\
636 & K1/K2III & 5.29 $\pm$ 0.01 & 1.03 $\pm$ 0.01 & -0.303 $\pm$ 0.019 & 12.60 $\pm$ 0.08 & 79.2 $\substack{+0.5 \\ -0.5}$ & 0.80 $\pm$ 0.02 & 1.177 $\pm$ 0.005 & 0.47 $\pm$ 0.02 & 4879 $\pm$ 51 & 2.78 $\pm$ 0.15 & 0.19 $\pm$ 0.04 & 2.23 $\pm$ 0.09 & 50.47 $\pm$ 1.17 & 9.94 $\pm$ 0.24 \\
770 & K0III & 6.54 $\pm$ 0.01 & 1.04 $\pm$ 0.02 & -0.317 $\pm$ 0.017 & 7.22 $\pm$ 0.04 & 137.9 $\substack{+0.7 \\ -0.7}$ & 0.84 $\pm$ 0.02 & 1.178 $\pm$ 0.003 & 0.54 $\pm$ 0.01 & 4845 $\pm$ 45 & 2.66 $\pm$ 0.10 & -0.08 $\pm$ 0.04 & 1.91 $\pm$ 0.17 & 49.14 $\pm$ 1.04 & 9.95 $\pm$ 0.21 \\
... & & & & ... & & & & ... & & & ... & & & & ...\\
224949 & K0III & 7.10 $\pm$ 0.01 & 0.99 $\pm$ 0.02 & -0.338 $\pm$ 0.013 & 5.73 $\pm$ 0.05 & 173.7 $\substack{+1.4 \\ -1.4}$ & 0.90 $\pm$ 0.02 & 1.183 $\pm$ 0.004 & 0.60 $\pm$ 0.02 & 4795 $\pm$ 32 & 2.49 $\pm$ 0.09 & -0.33 $\pm$ 0.03 & 1.30 $\pm$ 0.06 & 47.32 $\pm$ 1.07 & 9.97 $\pm$ 0.17 \\
\hline
\end{tabular}
}
\begin{tablenotes}
    \scriptsize
    \item {[1]} - HIPPARCOS catalog \citep{ESA1997}, [2] - TYCHO-2 catalog \citep{Hog2000}, [3] - \citet{Alonso1999}, [4] - GAIA DR2 \citep{Brown2018}, 
    
    [5] - \citet{Bailer-Jones2018}, [6] - this paper (see \autoref{sub:spec_param}), [7] - this paper, with evolutionary tracks from \citet{Pietrinferni2004}.
\end{tablenotes}
\end{center}
\label{tab:sample_table}
\end{table*}

% Table stellar parameters
\begin{table*}[!ht]
\centering
\begin{threeparttable}
\caption{Observed and inferred stellar parameters.}
%\begin{center}
    \begin{tabular}{lllccc}        
    \hline
    \hline
    & & ref. & HD\,22532 & HD\,64121 & HD\,69123\\
    TIC & & & 200851704 & 264770836 & 146264536\\
    GAIA DR2 & & & {\tiny 4832768399133598080} & {\tiny 5488303966125344512} &  {\tiny 5544699390684005248}\\
    \hline
    Sp. Type & & [1] & G8III/IV & G8/K0III & K1III \\
    $V$ & [mag] & [2] & 7.85 $\pm$ 0.01 & 7.44 $\pm$ 0.01 & 5.77 $\pm$ 0.01 \\
    $B-V$ & [mag] & [2] & 0.89 $\pm$ 0.02 & 0.86 $\pm$ 0.02 & 1.02 $\pm$ 0.01 \\
    $BC$ & & [3] & -0.250 $\pm$ 0.013 & -0.238 $\pm$ 0.012 & -0.318 $\pm$ 0.016 \\
    $\pi$ & [mas] & [4] & 6.18~$\pm$~0.03 & 7.67~$\pm$~0.03 & 13.28~$\pm$~0.06 \\
    $d$ & [pc] & [5] & 161.2 $\substack{+0.7 \\ -0.7}$ & 130.0 $\substack{+0.5 \\ -0.5}$ & 75.1 $\substack{+0.4 \\ -0.4}$ \\
    $M_{V}$ & [mag] & [2,4,5] & 1.81 $\pm$ 0.01 & 1.87 $\pm$ 0.01 & 1.39 $\pm$ 0.01 \\
    $Bp-Rp$ & [mag] & [4] & 1.087 $\pm$ 0.002 & 1.076 $\pm$ 0.004 & 1.183 $\pm$ 0.003 \\
    $M_G$ & [mag] & [4] & 1.56 $\pm$ 0.01 & 1.60 $\pm$ 0.01 & 1.09 $\pm$ 0.01 \\
    $T_{eff}$ & [K] & [4] & 5067 $\substack{+59 \\ -22}$ & 5066 $\substack{+58 \\ -60}$ & 4787 $\substack{+280 \\ -51}$ \\
    & & [6] & 5038 $\pm$ 24 & 5078 $\pm$ 22 & 4842 $\pm$ 41 \\
    $log\,g$ & [cm\,s$^{-2}$] & [6] & 3.09 $\pm$ 0.07 & 3.19 $\pm$ 0.06 & 2.86 $\pm$ 0.11 \\
    $[Fe/H]$ & [dex] & [6] & -0.19 $\pm$ 0.02 & -0.21 $\pm$ 0.02 & 0.05 $\pm$ 0.03 \\
    $M_{*}$ & [M$_{\odot}$] & [7] & 1.57 $\pm$ 0.07 & 1.64 $\pm$ 0.06 & 1.68 $\pm$ 0.09 \\
    & & [8] & 1.20 $\pm$ 0.05 & 1.18 $\pm$ 0.05 & 1.43 $\pm$ 0.07 \\
    $L_{*}$ & [L$_{\odot}$] & [2,3,4] & 18.80 $\pm$ 0.33 & 17.70 $\pm$ 0.30 & 29.51 $\pm$ 0.57 \\
    $R_{*}$ & [R$_{\odot}$] & [2,3,4,6] & 5.69 $\pm$ 0.07 & 5.44 $\pm$ 0.07 & 7.72 $\pm$ 0.15 \\
    \hline
    \end{tabular}
\begin{tablenotes}
    \small
    \item {[1]} - HIPPARCOS catalog \citep{ESA1997}, [2] - TYCHO-2 catalog \citep{Hog2000}, [3] - \citet{Alonso1999}, [4] - GAIA DR2 \citep{Brown2018}, [5] - \citet{Bailer-Jones2018}, [6] - this paper (see \autoref{sub:spec_param}), [7] - this paper, with evolutionary tracks from \citet{Pietrinferni2004}, [8] - \citet{Buldgen2021}.
\end{tablenotes}
%\end{center}
\label{tab:stellar_params}
\end{threeparttable}
\end{table*}

%--------------------------------------------------------------------
\section{Data acquisition and analysis}\label{sec:obs_analysis}

\subsection{Data acquisition and processing}\label{subsec:ts_obs}
For each target, we collected several tens of radial-velocity data over a median time span of 13 years, with a typical S/N = 70 for an exposure time between 180 and 300\,s\footnote{Following the 2007 and 2014 upgrades, we have to fit a small radial-velocity offset between the three versions of the CORALIE instrument, the values of the offset depending on several aspects such as the considered star or the correlation mask used. We thus consider the three versions of CORALIE as three different instruments.}. \cref{tab:timeseries_hd22532,tab:timeseries_hd64121,tab:timeseries_hd69123} give the list of these measurements with their instrumental error bars.
We first analyzed the radial-velocity time series using the radial-velocity module of the Data \& Analysis Center for Exoplanets (DACE) web platform,\footnote{\href{https://dace.unige.ch/radialVelocities/?}{https://dace.unige.ch/radialVelocities/?.}} which provides an open access to a wide range of exoplanets' observational and theoretical data with the corresponding data visualization and analysis tools. 
The formalism of the radial-velocity data analysis implemented in DACE is described in Ségransan et al. (submitted, Appendix~A) and is mainly based on algorithms presented in \citet{Diaz2014} and \citet{Delisle2016, Delisle2018}.

Our general approach for a periodic signal search is the following. For each time series, we follow an iterative process consisting of looking for successive significant dominant peaks in the periodograms of the corresponding radial-velocity residuals. At each step, the radial-velocity residuals are computed by readjusting the model composed of the N-independent Keplerians, potential linear or quadratic drift terms to fit long-term trends, the individual instrumental offsets, and additional noise. We fit a combination of white noise terms corresponding to individual instrumental precisions\footnote{The instrumental precisions are well constrained for each version of CORALIE, calibrated on non-active stars: $\sigma_{COR98}=5.0\pm0.5$\,m\,s$^{-1}$, $\sigma_{COR07}=8.0\pm0.5$\,m\,s$^{-1}$, and $\sigma_{COR14}=3.0\pm0.5$\,m\,s$^{-1}$. Those values are used as priors on the instrumental noise terms in \autoref{sec:results}.} and a global noise term attributed to intrinsic stellar jitter. This approach allows us to obtain an idea of how much noise can be attributed to stellar physics; however, one must be aware of the degeneracy between those two sources of noise, which is only partially lifted by using strong priors on the instrumental noise. The final error bars on the velocities correspond to the quadratic sum of the error computed by the data reduction software, the instrumental noise and the stellar jitter. We proceeded with the periodicity search by computing the periodogram of the data in the $1-10\,000$ days\footnote{Using the algorithm implemented on DACE (see \citet{Delisle2020b}) and setting the upper bound of the periodogram at approximately twice the time span of the survey.} range and using the false alarm probability (FAP) to assess the significance of the signal, following the formalism of \citet{Baluev2008}. Significant signals can have different origins, and they are discussed in \autoref{subsec:stell_lineprof}.

\subsection{Stellar activity and line profile analysis}\label{subsec:stell_lineprof}
Stellar activity in giant stars originates from different phenomena. Short period modulations of the order of hours to days (first discussed by \citet{Walker1989,Hatzes1993,Hatzes1994}) are understood to be the result of solar-like radial pulsations (p, g, or mixed modes) \citep{Frandsen2002,DeRidder2006,Hekker2006a}.
Concerning longer period variations, mechanisms such as magnetic cycles \citep{Santos2010,Dumusque2011a}, rotational modulation of features on the stellar surface (star spots, granulation, etc. \citep{Lambert1987,Larson1993,DelgadoMena2018}), beating of modes, or a combination of all three are to be considered. Non-radial oscillations have also been discussed \citep{Hekker2006c} and confirmed by  \citet{DeRidder2009,Hekker2010b} as a source of periodic modulation of the spectroscopic cross-correlation profile. Those modes can have lifetimes of up to several hundreds of days \citep{Dupret2009}.

The careful monitoring of the spectral line profile via the cross-correlation function (CCF) and of the chromospheric activity indicators is essential to help distinguish between planetary signals and stellar-induced variations of the radial velocities.

Our estimate of the star's radial velocity is based on the CCF technique \citep{Griffin1967,Baranne1979,Queloz2001}, which creates a sort of mean spectral line from the thousands of lines used in the correlation, and of significantly higher S/N compared to a single line. In order for stellar activity to significantly impact the CCF profile, and thus the radial-velocity value, it would have to affect the majority of the spectral lines. Such an effect could cause deformations in the line profile and possibly mimic a planetary signal. Computing the contrast, radial-velocity, full width at half maximum (FWHM), and the bisector inverse span (BIS), which are linked to the first four moments of the line profile, gives enough information to precisely control the evolution of the profile along the time series \citep{Aerts2000}.

Magnetic activity enhances the emission from the stellar corona and chromosphere, resulting in emissions in the X-Ray and UV regions, as well as emissions in the cores of the \textit{H\&K Ca II} lines and H$\alpha$. The H$\alpha$ activity index is sensitive to solar prominences and chromospheric activity. The reversal emission in the line core of \textit{Ca II H\&K} (S-index) \citep{Wilson1978}, which measures the contributions from the stellar photosphere and chromosphere, and the $log\,R'_{HK}$ activity index, which measures the chromospheric contribution of the \textit{H\&K Ca} lines excluding the photospheric component, cannot be directly computed from the CORALIE spectra in a reliable way, because of the low S/N in this part of the spectra.

The time series and corresponding periodograms of those line profiles and of the H$\alpha$ chromospheric indicator \citep[following the method described in][]{Boisse2009,GomesdaSilva2011} are produced systematically to check for any signs of periodicity and a possible origin of the radial-velocity variations.  Correlations between these indicators and radial velocities were also checked. Causes such as stellar pulsations can be ruled out by comparing the behavior of line profiles from different spectral regions; for pulsating stars, the temporal and phasing behavior of the moments should remain the same for any spectral region (a signature should also be present in the BIS \citep{Hatzes1999}). For detailed examples, we invite the reader to consult the analyses of \cite{Briquet2001b} or \citet{Briquet2004}, which attempted to discriminate between stellar pulsation and rotational modulation (presence of stellar spots) as the source of observed periodic variability, using \textit{Si II} and \textit{He I} lines in slow pulsating B stars. In the case of rotational modulation, the BIS and the S-index should vary in phase and with the same period as the radial-velocities, which is a period that should also correspond to the stellar rotation period. However, phase shifts have been observed, for example, in the case of the G2 dwarf HD\,41248 \citep{Santos2014}. The star exhibits a 25\,day periodicity in the radial-velocity, FWHM, and $log\,R'_{HK}$ time series, probably explained by rotational modulation combined with a strong differential rotation of the star.
We should, however, stress here that giant stars are still not fully understood, and we have to keep in mind that the absence of periodic signals in line-shape variations does not mean for sure that the radial-velocity signal is induced by a planetary companion. It remains, however, our best interpretation of the observations.

%--------------------------------------------------------------------
\section{Analysis of individual systems: Orbital solutions}\label{sec:results}

Following the approach described in \autoref{subsec:ts_obs}, we analyzed the long time series of observations obtained for the three targets presented in this paper. The final parameters of each system were computed using the MCMC algorithm implemented in DACE (developed by \citet{Diaz2014,Diaz2016}) to probe the complete parameter space, with $1.6$ million iterations. We fit the following parameters for the Keplerian model: log\,P and log\,K to better explore ranges of several orders of magnitude with a uniform prior. $\sqrt{e \cos{\omega}}$ and $\sqrt{e \sin{\omega}}$ to obtain a uniform prior for the eccentricity. Finally, we obtained $\lambda_0,$ the mean longitude at epoch of reference (i.e., $BJD=2\,455\,500$ [d]), with a uniform prior. We used a uniform prior for the COR07 offset of reference, and Gaussian priors for the relative offsets between COR07 and COR98/14: $\Delta\,RV_{COR98-COR07}$: $\mathcal{N}(0,4)$ m\,s$^{-1}$ and $\Delta\,RV_{COR14-COR07}$: $\mathcal{N}(14,4)$ m\,s$^{-1}$. We also used Gaussian priors for the instrumental noise: $\sigma_{COR98}$: $\mathcal{N}(5,1)$ m\,s$^{-1}$, $\sigma_{COR07}$: $\mathcal{N}(8,1.5)$ m\,s$^{-1 }$ , and $\sigma_{COR14}$: $\mathcal{N}(3,0.5)$ m\,s$^{-1}$. Finally, we used a uniform prior for the stellar jitter parameter.

For all three targets, the fit instrumental noises (on top of the photon noise) are in the range of the individual instrumental precisions. In the case of HD\,64121, the fit stellar jitter clearly dominates over the instrumental precision, with a level of $\sim$17\,m\,s$^{-1}$. For each time series, we checked for periodicities and correlations in the activity-related products of the 
high-resolution spectra mentioned above. 

\subsection{HD 22532}\label{subsec:hd22532}
\begin{figure}[!p]
        \centering
        \adjincludegraphics[width=1\columnwidth, trim={0 0 0 {-.004\height}},clip]{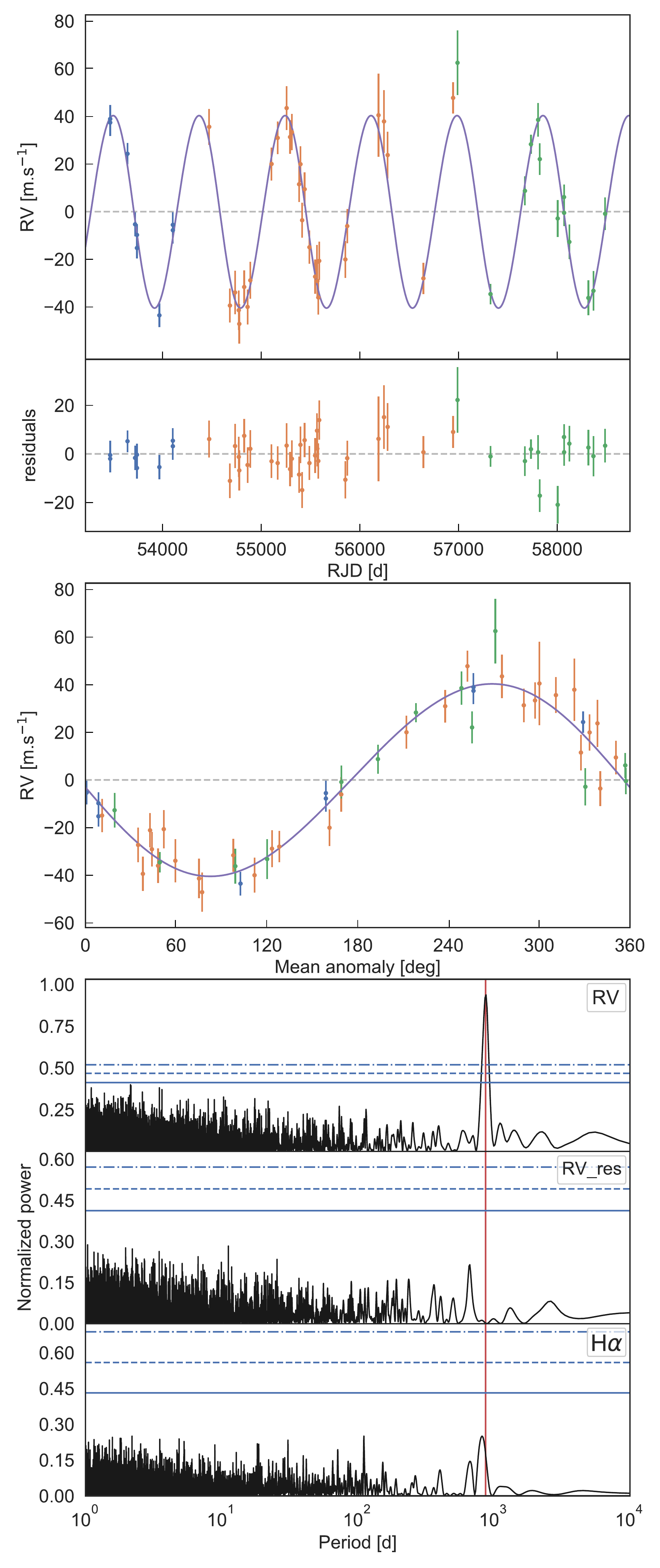}
        \caption{Top panel: Radial velocities of HD\,22532 from CORALIE (COR98 in blue, COR07 in orange and COR14 in green) with the best Keplerian model superimposed (solid line), and corresponding residuals around the solution. Second panel: Phased radial-velocity solution. Third panel: Periodograms of the radial-velocity time series, the residuals of the radial-velocities after substraction of the fit periodic signal, and the periodiodogram of the H$\alpha$ activity index time series. The red vertical line indicates the period of the orbital solution (872.6\,days). Horizontal lines are the FAP levels at 10\% (continuous), 1\% (dashed), and 0.1\% (dotted-dashed).}
        \label{fig:timeseries_hd22532}
\end{figure}

For HD\,22532\footnote{TIC\,200851704 ; GAIA DR2 4832768399133598080} , we detect a $\sim$873\,day periodic variation of the radial velocities, which, fit by a Keplerian model, corresponds to a planet in a quasi-circular orbit, 1.9\,au away from its star, and with a semi-amplitude of 40\,m\,s$^{-1}$ corresponding to a minimum mass of 2.1\,M$_J$ (using $M_{\star}=1.2$\,M$_{\odot}$ from \autoref{tab:stellar_params}). We observe in the periodogram of the H$\alpha$ activity index of HD\,22532 (see bottom periodogram in \autoref{fig:timeseries_hd22532}) a non-significant peak at $\sim$810 days, with a FAP level well below 10\%, which is at the same level as the higher frequency white noise. We checked for a linear correlation with the radial-velocities, and the weighted Pearson coefficient was found as low as $R_{P}=-0.396\pm0.068$. We also computed the weighted Spearman's rank $R_S=0.411\pm0.066$, which is also considered a low correlation. The important dispersion of the H$\alpha$ data points and the low significance of a period approximately 60 days shorter than the one detected in the radial velocities indicates that those variations are most likely not at the origin of - nor are they linked to - the radial-velocity signal.

\begin{figure}[!pt]
        \centering
        \adjincludegraphics[width=1\columnwidth, trim={0 0 0 0},clip]{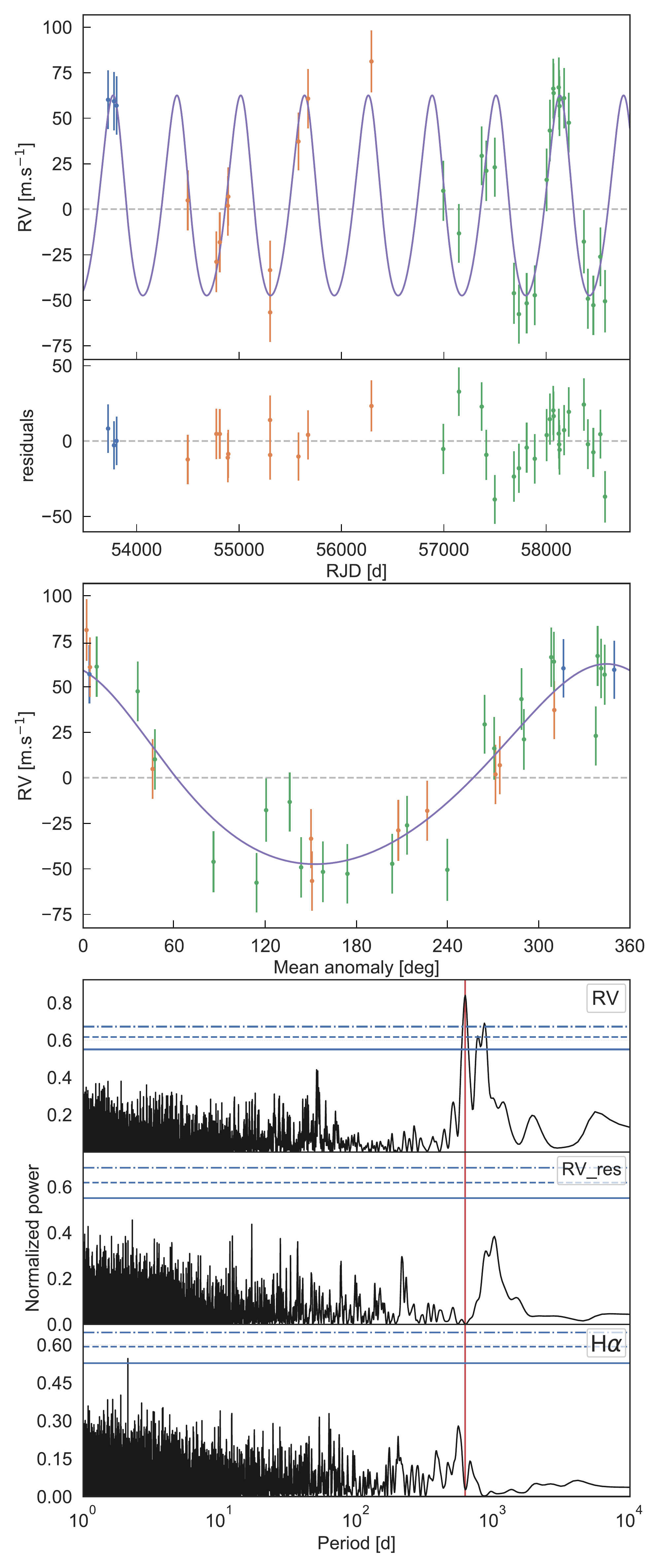}
        \caption{Same as \autoref{fig:timeseries_hd22532}, but for HD\,64121. The period of the best solution is 623.0\,days.}
        \label{fig:timeseries_hd64121}
\end{figure}

\begin{figure}[!pt]
        \centering
        \adjincludegraphics[width=1\columnwidth, trim={0 0 0 0},clip]{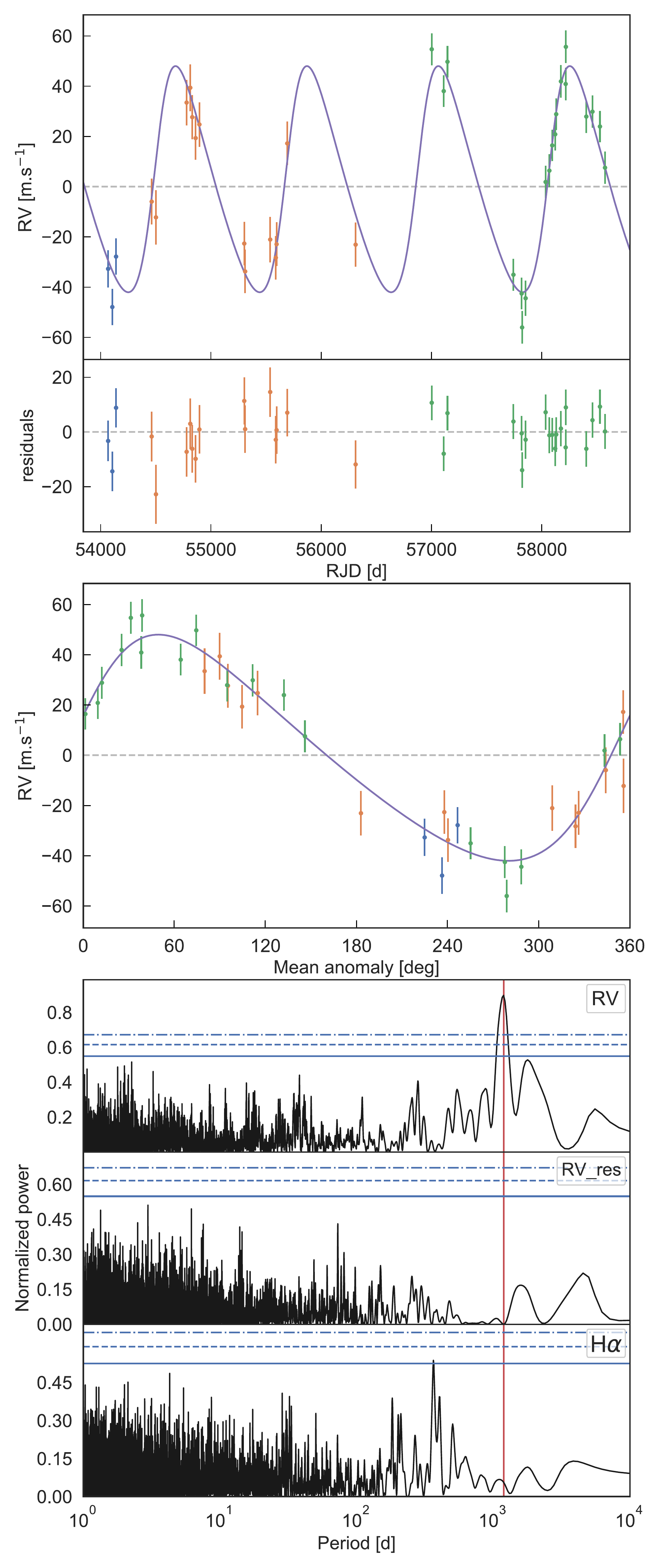}
        \caption{Same as \autoref{fig:timeseries_hd22532}, but for HD\,69123. The period of the best solution is 1193.3\,days.}
        \label{fig:timeseries_hd69123}
\end{figure}

\subsection{HD 64121}\label{subsec:hd64121}
In the case of HD\,64121\footnote{TIC\,264770836 ; GAIA DR2 5488303966125344512}, a $\sim$623 day periodic variation is fit by a Keplerian model. It corresponds to a planet in a low-eccentricity orbit, 1.5\,au away from its star, with a semi-amplitude of $\sim$55\,m\,s$^{-1}$ and corresponding to a minimum mass of 2.6\,M$_J$ ($M_{\star}=1.18$\,M$_{\odot}$). The periodogram of the radial-velocity residuals, after substraction of the fit, presents a non-significant peak at $\sim$1\,000\,days at the same level as the higher-frequency noise. HD\,64121 also exhibits a similar non-significant peak in the periodogram of the H$\alpha$ activity index (see bottom periodogram in \autoref{fig:timeseries_hd64121}), at $\sim$550 days. The weighted Pearson correlation coefficient with the radial-velocities was found to be non-significant at $R_P=0.072\pm0.116$. We also computed the weighted Spearman's rank $R_S=0.100\pm0.118$, which is also non-significant. We reach the same conclusion as for HD\,22532, that those variations are most likely not at the origin of - nor are they linked to - the radial-velocity signal.

\subsection{HD 69123}\label{subsec:hd69123}
Finally, HD\,69123\footnote{TIC\,146264536; GAIA DR2 5544699390684005248.} presents the longest periodic variation, with a $\sim$1193 day signal corresponding to a planet in a slightly eccentric orbit of $e=0.2$. The semi-major axis of the planetary orbit is $\sim$2.5\,au, and the semi-amplitude $\sim$47\,m\,s$^{-1}$ leads to a minimum mass of 3\,M$_J$ for the planetary companion ($M_{\star}=1.43$\,M$_{\odot}$). HD\,69123 presents a peak in the periodogram of the H$\alpha$ activity index, with a FAP level below 1\% (bottom panel in \autoref{fig:timeseries_hd69123}), at a period of $\sim$367 days. We suspect this almost one-year periodicity to be caused by a telluric contamination of the H$\alpha$ line, and potentially water lines.

\subsection{Intrinsic variability and final solutions}

For the three stars, none of our other activity indicators (contrast, FWHM and BIS) show any similar periodicity or significant correlation with the radial velocities (see \autoref{apdx:perio_act_indic}).

%%%%%%%%%%%%
% Photometry
%%%%%%%%%%%%
We also checked the V-band photometric data available in the All-Sky Automated Survey (ASAS-3, \citet{Pojmanski2002}) for our stars. This survey is very interesting as it is one of the only surveys with a time span of almost nine years. For reasons of consistency and reliability of the data post-processing (mainly correction of saturation and camera focus stability due to instrumental issues), we have to consider this data with caution when using it to check for variability due to intrinsic stellar processes or surface rotational modulation. We discuss this matter in more detail in \citet{Pezzotti2021}. No periodicities linked to the ones detected in the radial-velocity data have been found for any of the three stars presented in this paper.

% Table orbital parameters
\begin{table*}[!htb]
\centering
\begin{threeparttable}
\caption{Radial-velocity observation statistics, best-fit solutions of the model with instrumental offsets, nuisance parameters, Keplerian orbital parameters, and inferred planetary parameters.}
%\begin{center}
    \begin{tabular}{llccc}        
    \hline
    \hline
    & & HD\,22532b & HD\,64121b & HD\,69123b \\
    \hline
    \multicolumn{5}{c}{\textbf{Observations}}\\
    \hline
    $N_{obs}$ & & 52 & 36 & 36 \\
    $T_{span}$ & $[days]$ & 5016 & 4853 & 4507 \\
    $rms_{tot}$ & $[m.s^{-1}]$ & 31.15 & 44.93 & 31.68 \\
    $rms_{res}$ & $[m.s^{-1}]$ & 8.44 & 15.93 & 8.42 \\
    $\chi^2_{red}$ & & 1.30 & 1.44 & 1.69 \\
    \hline
    \multicolumn{5}{c}{\textbf{Offsets $^{(1)}$}}\\
    \hline
    $\Delta\,RV_{COR98-COR07}$ & $[m/s]$ & 2.0~$\pm$~2.3 & -0.1~$\pm$~3.7 & -4.1~$\pm$~3.5 \\
    $\Delta\,RV_{COR07-COR07}$ & $[m/s]$ & 29248.9~$\pm$~1.5 & -4117.9~$\pm$~4.0 & 27476.7~$\pm$~2.4 \\
    $\Delta\,RV_{COR14-COR07}$ & $[m/s]$ & 20.9~$\pm$~2.2 & 15.1~$\pm$~3.4 & 21.9~$\pm$~2.8 \\
    \hline
    \multicolumn{5}{c}{\textbf{Instrumental Noises}}\\
    \hline
    $\sigma_{COR98}$ & $[m/s]$ & 4.7~$\pm$~1.0 & 4.9~$\pm$~1.0 & 5.2~$\pm$~1.0 \\
    $\sigma_{COR07}$ & $[m/s]$ & 6.8~$\pm$~1.2 & 7.8~$\pm$~1.5 & 7.8~$\pm$~1.4 \\
    $\sigma_{COR14}$ & $[m/s]$ & 3.1~$\pm$~0.5 & 3.0~$\pm$~0.5 & 3.0~$\pm$~0.5 \\
    \hline
    \multicolumn{5}{c}{\textbf{Stellar Jitter}}\\
    \hline
    $\sigma_{jit}$ & $[m.s^{-1}]$ & 2.1~$\pm$~1.6 & 16.8~$\pm$~2.6 & 7.2~$\pm$~1.8 \\
    \hline
    \multicolumn{5}{c}{\textbf{Keplerians}}\\
    \hline
    $P$ & $[days]$ & 872.6~$\pm$~2.8 & 623.0~$\pm$~3.4 & 1193.3~$\pm$~7.0 \\
    $K$ & $[m.s^{-1}]$ & 40.0~$\pm$~1.6 & 55.2~$\pm$~4.1 & 46.8~$\pm$~2.4 \\
    $e$ & & 0.03~$\pm$~0.03 & 0.11~$\pm$~0.07 & 0.19~$\pm$~0.06 \\
    $\omega$ & $[deg]$ & 169.1~$\pm$~88.7 & 2.7~$\pm$~56.0 & -67.3~$\pm$~21.7 \\
    $\lambda_0$ $^{(2)}$ & $[deg]$ & 110.7~$\pm$~2.3 & -77.5~$\pm$~7.3 & 227.6~$\pm$~4.5 \\
    $T_p$ $^{(2)}$ & $[rjd]$ & 5575.0~$\pm$~221.0 & 5653.0~$\pm$~130.0 & 5715.7~$\pm$~64.6 \\
    \hline
    $a$ & $[au]$ & 1.900~$\pm$~0.004 & 1.510~$\pm$~0.006 & 2.482~$\pm$~0.010 \\
    $m_2\,sin\,i$ $^{(3)}$ & $[M_J]$ & 2.12~$\pm$~0.09 & 2.56~$\pm$~0.19 & 3.04~$\pm$~0.16 \\    
    \hline
    \end{tabular}
\begin{tablenotes}
    \small 
    \item {$^1$} The reference instrument is COR07.
    \item {$^2$} The mean longitude is given at $BJD=2\,455\,500$ [d], while $2\,450\,000$ has been subtracted from the date of passage through periastron ($T_p$).
    \item {$^3$} Using the model-independent mass from seismic inversions \citep[see][]{Buldgen2021}).\end{tablenotes}
%\end{center}
\label{tab:orbit_params}
\end{threeparttable}
\end{table*}

%%%%%%%%%%%%
% Conclusion
%%%%%%%%%%%%
We are thus fairly confident that the observed radial-velocity periodic variations are not due to chromospheric stellar activity, or rotational modulation of surface features such as spots, which would require an significant percentage of the stellar surface to be covered. We also found no indication of long-period, non-radial oscillation modes (neither matching periodicities nor corresponding harmonics in the line profile moments). We thus consider that the observed radial-velocity signals are due to planetary companions orbiting the stars. The resulting models and residuals are shown in \cref{fig:timeseries_hd22532,fig:timeseries_hd64121,fig:timeseries_hd69123}, overplotted on the radial-velocities. In \autoref{tab:orbit_params}, we present the statistics of the distributions (i.e., the median and standard deviation) of the more common set of Keplerian parameters P, K, e, $\omega$ and T$_P$, as well as the distributions of the semi-major axis and minimum masses, derived from the MCMC chains of the fit parameters. In \autoref{apdx:corner}, we present the corner plots of the posterior distributions of the fit parameters for each star. The weighted rms scatter of the radial velocities $rms_{tot}$ and residuals to the Keplerian fit $rms_{res}$ are also provided in the table. For all three targets, the rms of the residuals are comparable to those of single giant stars with similar $B-V$ \citep[see][Fig.\,3]{Hekker2006b}.

%--------------------------------------------------------------------
\section{Discussion and conclusion}\label{sec:conclusion}

Since 2006, we have been conducting a high-precision radial-velocity survey of a volume-limited sample of 641 giant stars using the CORALIE spectrograph on the 1.2\,m Leonard Euler Swiss telescope at La Silla Observatory (Chile). Our goal is to better understand the formation and evolution of planets around stars more massive than the Sun, including the evolutionary stage of stars toward the giant branch, through a statistical study of the properties of the detected planet population. The sample is volume limited, targeting giant stars in the southern hemisphere (declination $<25^{\circ}$) inside a 300\,pc radius around the Sun. The evolutionary stage of the stars was constrained by magnitude ($M_V<2.5$) and color ($0.78<B-V<1.06$) cut-offs to avoid main-sequence stars and intrinsically variable late-type giants. We derived reliable spectroscopic parameters from CORALIE-14 spectra \citep[following][]{Santos2004,Sousa2014,Alves2015}. Our sample shows a distribution of a metallicity ratio of iron similar to the one of stars in the solar neighborhood, peaking between 0.0 and 0.1\,dex, but missing very low and rich metallicity stars. This may be explained by the young age of the giants, compared to their dwarf counterparts, which did not leave them enough time to migrate in the Galaxy \citet{Wang2013,Minchev2013}. A color cut-off bias could also be part of the explanation for this effect, excluding the low-log-g stars with high metallicity and the high-log-g stars with low metallicity, as discussed in \citet{Mortier2013, Adibekyan2019}. We also obtained stellar masses for the sample with global parameters fitting, using evolutionary tracks from \citet{Pietrinferni2004}, using the SPInS software \citep{Lebreton2020}. The distribution ranges approximately from 0.75 to 4\,M$_{\odot}$, with a maximum around 2\,M$_{\odot}$.

This paper is the first of a series in which we present the first results of the survey, namely the detection and characterization of three new planetary companions orbiting the giant stars HD\,22532, HD\,64121, and HD\,69123, taking advantage of asteroseismic masses, following the methodology of \citet{Buldgen2019}, obtained with the TESS data \citep{Ricker2014}. For each star, we systematically checked for any correlation with chromospheric activity, rotational modulation of surface features, or long-term non-radial pulsations. We also consulted the corresponding ASAS-3 photometry time series \citep{Pojmanski2002}, spanning 6.8 to 7.4 years. No significant periodicities or correlations linked to the radial-velocity signal detected have been found.

\begin{figure*}[t!]
        \centering
        \adjincludegraphics[width=1\textwidth, trim={0 0 0 0},clip]{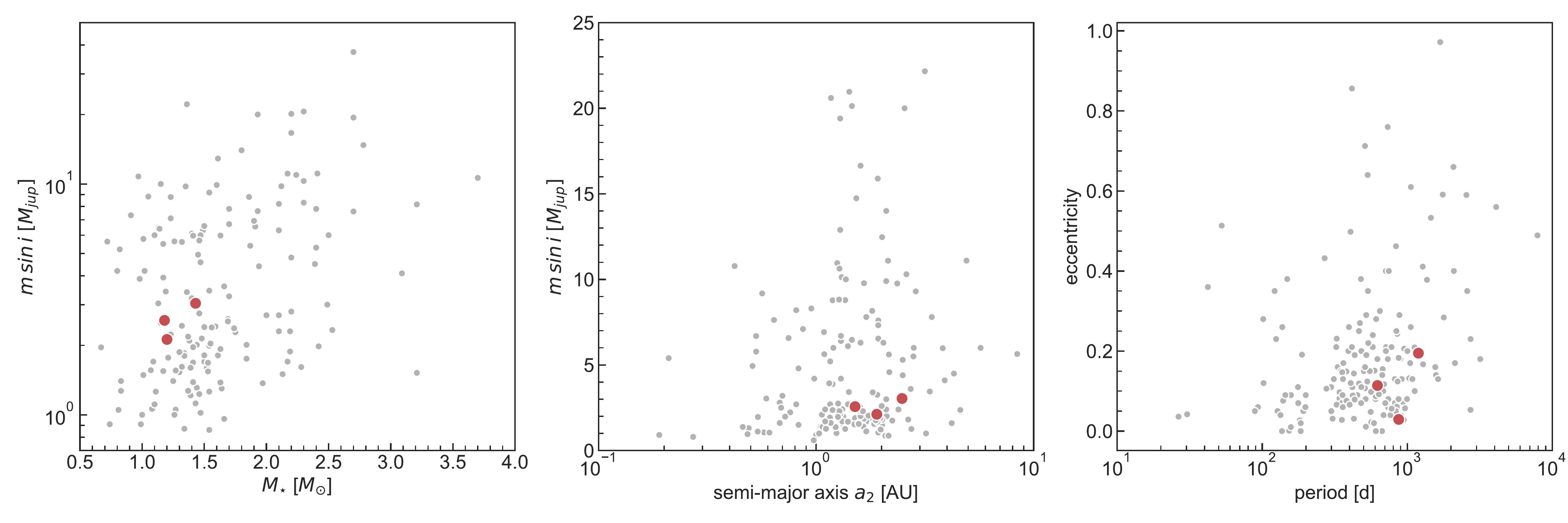}
        \caption{Stellar and planetary parameter relations for the 186 discovered planets orbiting giant stars. The three planets presented in this paper are represented by red dots.}
        \label{fig:pla_params}
\end{figure*}

The new planets are typical representatives of the known population of planets around giant stars, considering their masses, semi-major axes, and low eccentricities. This is illustrated in \cref{fig:pla_params} displaying the new detections together with the planet candidates from the literature. Most planets discovered around giant stars have eccentricities below 0.2-0.3 \citep[e.g.,][]{Jones2014,Yilmaz2017}, and are at distances, that is, farther away than 1\,au from the central star. 

%Forthcoming publications
Monitoring of the CASCADES sample is continuing. As an interesting by-product, it is also bringing important information on stellar binaries and star-brown-dwarf systems. The formation scenario for the latter is still unclear. The system may form initially as a binary star with an extreme mass ratio, or through a formation process comparable to the one for planets in the proto-stellar disk, via disk instability or core accretion. The maximum mass of planets forming in a massive disk is not known.  Forthcoming papers will present additional planetary candidates, as well as potential brown dwarfs and spectroscopic binaries found in the sample. We will also address through a statistical study of our sample, the main open questions linked with the planet population orbiting intermediate-mass (evolved) stars: distribution of orbital properties as constraints for planet formation models, correlations between planet characteristics and occurrence rate with primary star properties such as mass and metallicity.

In this context, some planet-host stars from our sample are particularly well suited for a deep asteroseismic analysis, giving access to their internal structure. The available information includes well-constrained planetary signals, long, photometric, high-precision time series from high-cadence observations with TESS, and accurate spectroscopic parameters. Asteroseismologic analysis can provide precious information concerning the past and future evolution of such systems. Among the highly interesting related questions is the one of the impact of stellar evolution on the planet orbits and of the potential engulfment of planets by the star \citep{Pezzotti2021}, for instance to explain the apparent lack of close-in, short-period planets ($P\leqslant100$\,days, $a\leqslant0.5$\,au). The second and third publications of the CASCADES series will focus on the asteroseimic analysis of the three stellar hosts presented in this paper \citep{Buldgen2021} and the analysis of a new planet-host star \citep{Pezzotti2021}, for which the full evolution of the system can be modeled.

% Interesting targets for transits
Giant stars hosting planets are good candidates for planetary transit searches. Due to the increase in radius at the giant stage, companions of giant stars have a higher probability of transit than planets around main-sequence stars. However, as these planets are on long-period orbits, the transit duration is on the order of tens to hundreds of hours. Moreover, because of the relative size of the planet and the star, the expected transit depth is very small. For our sample stars, they are in the $10-1000$ ppm range. Although very limiting for ground-based observations, these two aspects are tractable from space with satellites such as TESS \citep{Ricker2014} and CHEOPS \citep{Benz2020}. The three systems described in this paper present transit probabilities around 1.5\,\% and transit depth between 170 and 350\,ppm (considering planets with a 1\,R$_J$ radius). Unfortunately, none of these candidates has thus far had a transit time prediction in the window of the TESS observations.

%--------------------------------------------------------------------
\begin{acknowledgements}
We thank all observers at La Silla Observatory from the past fourteen years for their contribution to the observations and the quality of their work. We acknowledge financial support from the Swiss National Science Foundation (SNSF) for the project 2020-178930. This work has, in part, also been carried out within the framework of the National Centre for Competence in Research PlanetS supported by SNSF. In particular, this publication makes use of the The Data \& Analysis Center for Exoplanets (DACE, https://dace.unige.ch), a platform of Planets, based at the University of Geneva (CH), dedicated to extrasolar planet data visualisation, exchange and analysis. 
G.B. acknowledges funding from the SNF AMBIZIONE grant No. 185805 (Seismic inversions and modelling of transport processes in stars).
P.E. has received funding from the European Research Council (ERC) under the European Union’s Horizon 2020 research and innovation programme (grant agreement No 833925, project STAREX). 
C.P. acknowledges funding from the Swiss National Science Foundation (project Interacting Stars, number 200020-172505). 
V. A. acknowledges the support from FCT through Investigador FCT contract no. IF/00650/2015/CP1273/CT0001. 
N.C.S acknowledges support from FCT - Fundação para a Ciência e a Tecnologia through national funds and by FEDER through COMPETE2020 - Programa Operacional Competitividade e Internacionalização by these grants: UID/FIS/04434/2019; UIDB/04434/2020; UIDP/04434/2020; PTDC/FIS-AST/32113/2017 \& POCI-01-0145-FEDER-032113; PTDC/FIS-AST/28953/2017 \& POCI-01-0145-FEDER-028953. 
S.G.S acknowledges the support from FCT through Investigador FCT contract nr. CEECIND/00826/2018 and POPH/FSE (EC).
N.L. acknowledges financial support from "Programme National de Physique Stellaire" (PNPS) of CNRS/INSU, France.
This research has made use of the NASA Exoplanet Archive, which is operated by the California Institute of Technology, under contract with the National Aeronautics and Space Administration under the Exoplanet Exploration Program. 
\end{acknowledgements}

\bibliographystyle{aa} % style aa.bst
\bibliography{library} 

\begin{appendix}

\section{Online material - Radial-velocity data }\label{apdx:rv_timeseries}

% Table RV timeseries HD22532
\begin{table}[!hb]
\centering
\caption{Radial-velocity measurements and uncertainties for HD\,22532 obtained with the CORALIE spectrograph.}
\begin{threeparttable}
%\begin{center}
    \begin{tabular}{llll}        
    \hline
    \hline
    JD-2\,400\,000 & RV & e\_RV & Instrument\\
    & $[m\,s^{-1}]$ & $[m\,s^{-1}]$ & \\
    53469.478311 & 29290.32 & 5.11 & CORALIE98\\
53469.489712 & 29288.82 & 4.74 & CORALIE98\\
53644.880135 & 29275.72 & 3.38 & CORALIE98\\
53721.640656 & 29246.08 & 4.06 & CORALIE98\\
53740.549274 & 29241.64 & 3.52 & CORALIE98\\
53740.560662 & 29236.16 & 3.25 & CORALIE98\\
53967.913052 & 29207.90 & 4.05 & CORALIE98\\
54104.541987 & 29245.90 & 4.31 & CORALIE98\\
54104.553387 & 29243.63 & 4.70 & CORALIE98\\
    \hline
54472.657418 & 29283.05 & 4.61 & CORALIE07\\
54682.939662 & 29208.02 & 3.93 & CORALIE07\\
54734.882351 & 29213.52 & 6.86 & CORALIE07\\
54772.782614 & 29206.07 & 5.79 & CORALIE07\\
54777.726842 & 29200.32 & 5.71 & CORALIE07\\
54827.668046 & 29215.82 & 3.52 & CORALIE07\\
54861.626764 & 29207.45 & 4.26 & CORALIE07\\
54889.547083 & 29218.61 & 4.95 & CORALIE07\\
55104.768278 & 29267.47 & 3.49 & CORALIE07\\
55166.737312 & 29278.42 & 3.31 & CORALIE07\\
55257.569019 & 29290.95 & 6.97 & CORALIE07\\
55292.519484 & 29278.81 & 3.76 & CORALIE07\\
55310.462662 & 29280.85 & 4.70 & CORALIE07\\
55383.942691 & 29258.97 & 4.50 & CORALIE07\\
55397.859121 & 29267.41 & 4.65 & CORALIE07\\
55414.898447 & 29243.86 & 4.30 & CORALIE07\\
55439.891366 & 29256.90 & 3.76 & CORALIE07\\
55488.839456 & 29232.54 & 3.61 & CORALIE07\\
55546.594910 & 29220.18 & 4.09 & CORALIE07\\
55565.675911 & 29226.35 & 3.99 & CORALIE07\\
55568.618415 & 29218.36 & 4.09 & CORALIE07\\
55578.535688 & 29211.46 & 4.04 & CORALIE07\\
55587.635405 & 29226.81 & 5.37 & CORALIE07\\
55852.849665 & 29227.43 & 4.86 & CORALIE07\\
55871.644660 & 29241.42 & 4.03 & CORALIE07\\
56188.802495 & 29288.00 & 16.42 & CORALIE07\\
56244.610195 & 29285.39 & 11.70 & CORALIE07\\
56281.658388 & 29271.27 & 7.84 & CORALIE07\\
56643.747865 & 29219.51 & 2.73 & CORALIE07\\
56944.861399 & 29295.30 & 2.63 & CORALIE07\\
    \hline
56989.533778 & 29334.76 & 13.20 & CORALIE14\\
57323.731544 & 29237.66 & 2.97 & CORALIE14\\
57672.841328 & 29281.02 & 5.27 & CORALIE14\\
57733.669500 & 29300.58 & 2.59 & CORALIE14\\
57806.516799 & 29310.84 & 6.31 & CORALIE14\\
57823.510176 & 29294.34 & 5.93 & CORALIE14\\
58004.841570 & 29269.43 & 7.14 & CORALIE14\\
58068.689478 & 29278.41 & 4.27 & CORALIE14\\
58069.754285 & 29271.81 & 4.60 & CORALIE14\\
58122.674719 & 29259.59 & 6.61 & CORALIE14\\
58315.907826 & 29236.10 & 6.68 & CORALIE14\\
58366.842150 & 29239.08 & 7.75 & CORALIE14\\
58485.581907 & 29271.44 & 6.19 & CORALIE14\\
    \hline
    \end{tabular}
\begin{tablenotes}
    \footnotesize
    \item We note that small radial-velocity offsets between each instruments have to be considered. The offsets between COR98 and COR07 and between COR14 and COR07 are considered as free parameters in the model (see \autoref{tab:orbit_params}).
\end{tablenotes}
%\end{center}
\end{threeparttable}
\label{tab:timeseries_hd22532}
\end{table}

% Table RV timeseries HD64121
\begin{table}[!ht]
\centering
\caption{Radial-velocity measurements and uncertainties for HD\,64121 obtained with the CORALIE spectrograph.}
\begin{threeparttable}
%\begin{center}
    \begin{tabular}{llll}        
    \hline
    \hline
    JD-2\,400\,000 & RV & e\_RV & Instrument\\
    & $[m\,s^{-1}]$ & $[m\,s^{-1}]$ & \\
    53720.804034 & -4060.69 & 2.91 & CORALIE98\\
53778.680612 & -4061.51 & 2.56 & CORALIE98\\
53803.605854 & -4063.90 & 2.76 & CORALIE98\\
    \hline
54498.738216 & -4115.92 & 5.18 & CORALIE07\\
54778.828477 & -4149.64 & 6.10 & CORALIE07\\
54811.805888 & -4138.88 & 5.41 & CORALIE07\\
54889.595806 & -4118.86 & 4.85 & CORALIE07\\
54894.556246 & -4113.83 & 3.58 & CORALIE07\\
55302.504568 & -4154.21 & 4.40 & CORALIE07\\
55303.624298 & -4177.41 & 4.61 & CORALIE07\\
55579.781780 & -4083.50 & 3.20 & CORALIE07\\
55673.598495 & -4059.93 & 4.68 & CORALIE07\\
56292.793971 & -4039.50 & 6.72 & CORALIE07\\
    \hline
56993.853501 & -4092.41 & 4.82 & CORALIE14\\
57147.557314 & -4115.81 & 3.22 & CORALIE14\\
57369.806710 & -4073.18 & 2.93 & CORALIE14\\
57414.644584 & -4081.40 & 5.05 & CORALIE14\\
57496.490076 & -4079.47 & 3.37 & CORALIE14\\
57683.819281 & -4148.65 & 5.41 & CORALIE14\\
57732.846915 & -4160.13 & 3.66 & CORALIE14\\
57808.614926 & -4154.18 & 4.96 & CORALIE14\\
57887.545841 & -4149.76 & 4.23 & CORALIE14\\
58003.902541 & -4086.34 & 6.80 & CORALIE14\\
58034.854274 & -4059.25 & 6.03 & CORALIE14\\
58068.819513 & -4036.22 & 3.90 & CORALIE14\\
58071.860251 & -4038.67 & 3.90 & CORALIE14\\
58121.684266 & -4035.53 & 4.36 & CORALIE14\\
58125.767030 & -4042.36 & 3.95 & CORALIE14\\
58129.671706 & -4045.81 & 4.37 & CORALIE14\\
58173.764097 & -4041.42 & 4.82 & CORALIE14\\
58220.512518 & -4054.95 & 4.19 & CORALIE14\\
58366.910528 & -4120.28 & 7.19 & CORALIE14\\
58406.853503 & -4151.65 & 4.67 & CORALIE14\\
58459.644905 & -4155.20 & 3.85 & CORALIE14\\
58527.581933 & -4128.58 & 3.14 & CORALIE14\\
58573.499140 & -4153.03 & 6.36 & CORALIE14\\
    \hline
    \end{tabular}
\begin{tablenotes}
    \footnotesize
    \item Same as \autoref{tab:timeseries_hd22532}.
\end{tablenotes}
%\end{center}
\end{threeparttable}
\label{tab:timeseries_hd64121}
\end{table}

% Table RV timeseries HD69123
\begin{table}[ht]
\centering
\caption{Radial-velocity measurements and uncertainties for HD\,69123 obtained with the CORALIE spectrograph.}
\begin{threeparttable}
%\begin{center}
    \begin{tabular}{llll}        
    \hline
    \hline
    JD-2\,400\,000 & RV & e\_RV & Instrument\\
    & $[m\,s^{-1}]$ & $[m\,s^{-1}]$ & \\
    54066.795409 & 27431.94 & 2.21 & CORALIE98\\
54104.739993 & 27416.74 & 1.63 & CORALIE98\\
54138.713749 & 27436.81 & 1.42 & CORALIE98\\
    \hline
54461.826098 & 27471.03 & 3.81 & CORALIE07\\
54500.653028 & 27464.76 & 6.99 & CORALIE07\\
54778.840694 & 27510.46 & 3.66 & CORALIE07\\
54811.830340 & 27516.37 & 4.22 & CORALIE07\\
54829.726766 & 27504.65 & 2.92 & CORALIE07\\
54860.593499 & 27496.33 & 2.65 & CORALIE07\\
54894.597395 & 27501.75 & 3.19 & CORALIE07\\
55301.540504 & 27454.35 & 2.54 & CORALIE07\\
55309.605268 & 27443.25 & 2.62 & CORALIE07\\
55536.858310 & 27455.94 & 3.65 & CORALIE07\\
55587.747458 & 27448.72 & 2.54 & CORALIE07\\
55594.692659 & 27454.08 & 2.76 & CORALIE07\\
55691.556206 & 27494.23 & 2.62 & CORALIE07\\
56311.717839 & 27453.93 & 3.06 & CORALIE07\\
    \hline
57001.851834 & 27555.54 & 2.52 & CORALIE14\\
57110.683730 & 27538.88 & 2.44 & CORALIE14\\
57144.565060 & 27550.56 & 2.37 & CORALIE14\\
57742.816909 & 27465.79 & 2.65 & CORALIE14\\
57817.519883 & 27458.31 & 2.45 & CORALIE14\\
57821.536651 & 27444.81 & 2.90 & CORALIE14\\
57853.479450 & 27456.43 & 3.84 & CORALIE14\\
58034.879487 & 27502.73 & 2.77 & CORALIE14\\
58068.785220 & 27507.23 & 2.95 & CORALIE14\\
58094.717728 & 27517.27 & 2.24 & CORALIE14\\
58121.688868 & 27521.69 & 2.72 & CORALIE14\\
58130.724735 & 27529.70 & 2.48 & CORALIE14\\
58173.733307 & 27542.76 & 2.76 & CORALIE14\\
58216.553852 & 27541.73 & 2.90 & CORALIE14\\
58218.506427 & 27556.51 & 2.87 & CORALIE14\\
58403.872417 & 27528.76 & 2.91 & CORALIE14\\
58459.660012 & 27530.71 & 2.75 & CORALIE14\\
58527.594463 & 27524.80 & 2.32 & CORALIE14\\
58573.495729 & 27508.41 & 2.67 & CORALIE14\\
    \hline
    \end{tabular}
\begin{tablenotes}
    \footnotesize
    \item Same as \autoref{tab:timeseries_hd22532}.
\end{tablenotes}
%\end{center}
\end{threeparttable}
\label{tab:timeseries_hd69123}
\end{table}

\clearpage

\twocolumn[{\section{MCMC - corner plots distributions of fit parameters}\label{apdx:corner}}]
 
\begin{figure}[!h]
\centering
\begin{center}
\adjincludegraphics[width=\textwidth, trim={0 0 0 0},clip]{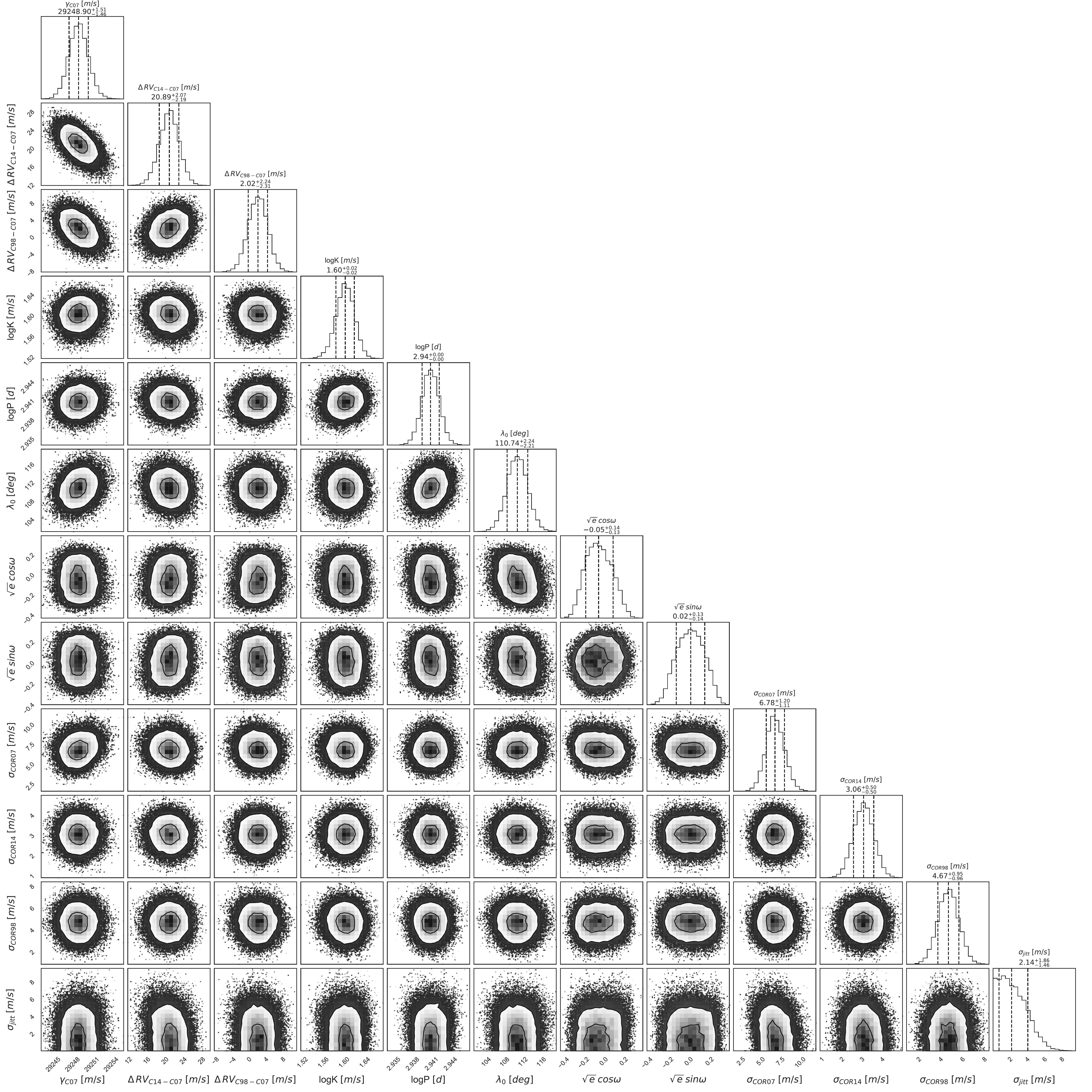}\\
\adjustbox{minipage=\textwidth,left}{\caption{Posterior distributions of fit parameters of HD22532. Each panel contains the two-dimensional histograms of the 1\,200\,000 samples (after removal of the burn-in, the first 25\% of the chains). Contours are drawn to improve the visualization of the 1$\sigma$ and 2$\sigma$ confidence interval levels. The upper panels of the corner plot show the probability density distributions of each orbital parameter of the final MCMC sample. The vertical dashed lines mark the 16th, 50th, and 84th percentiles of the overall MCMC samples, delimiting the 1$\sigma$ confidence interval.}}
\label{fig:hd22532_corner}
\end{center}
\end{figure}
% }]

\begin{figure*}[ht]
        \centering
        \adjincludegraphics[width=\textwidth, trim={0 0 0 0},clip]{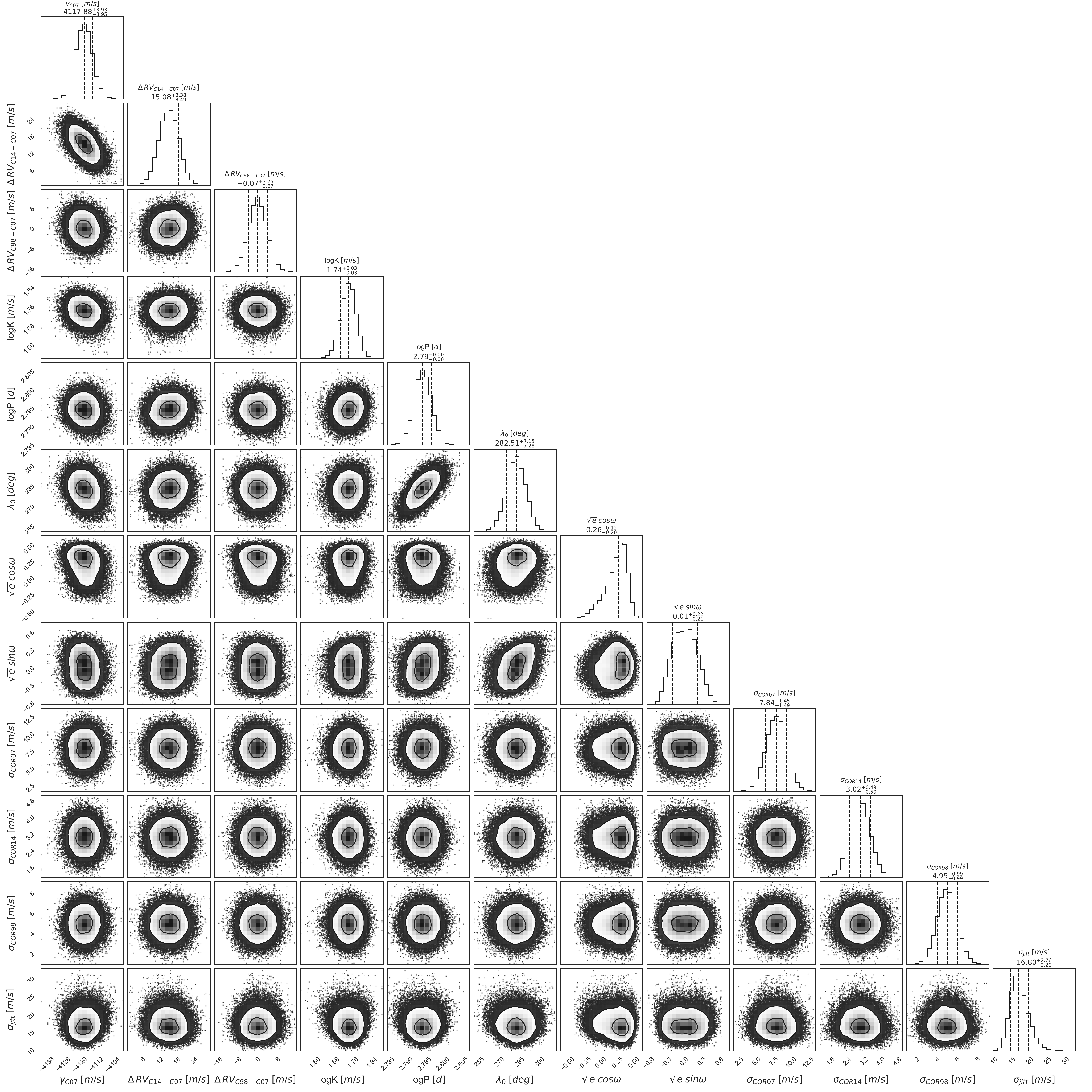}\\
        \caption{Same as \autoref{fig:hd22532_corner}, but for HD\,64121.}
        \label{fig:hd64121_corner}
\end{figure*}

\begin{figure*}[ht]
        \centering
        \adjincludegraphics[width=\textwidth, trim={0 0 0 0},clip]{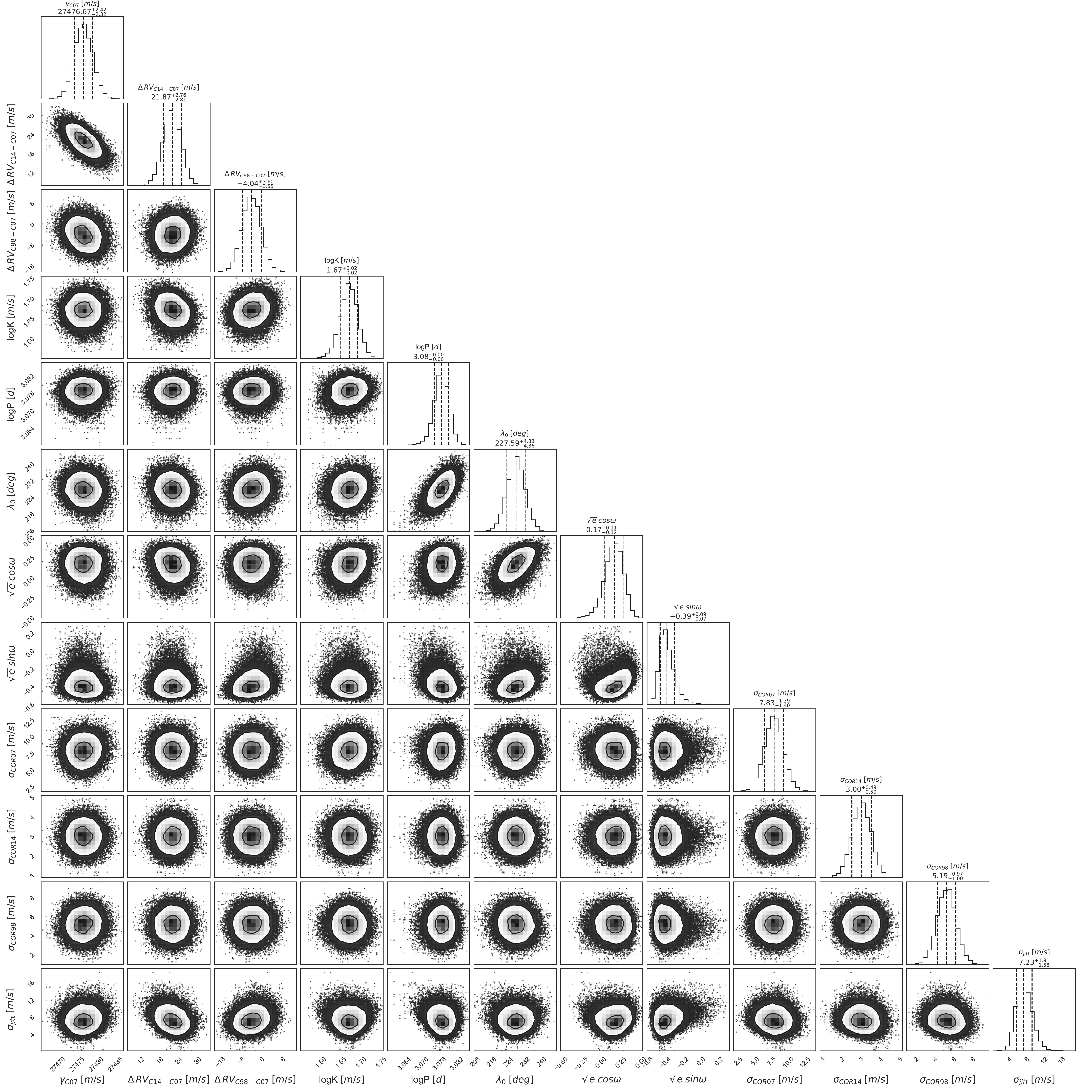}\\
        \caption{Same as \autoref{fig:hd22532_corner}, but for HD\,69123.}
        \label{fig:hd69123_corner}
\end{figure*}

\clearpage

\section{Intrinsic variability analysis - periodograms of line profile indicators}\label{apdx:perio_act_indic}

\begin{figure}[!ht]
        \centering
        \adjincludegraphics[width=\columnwidth, trim={0 0 0 0},clip]{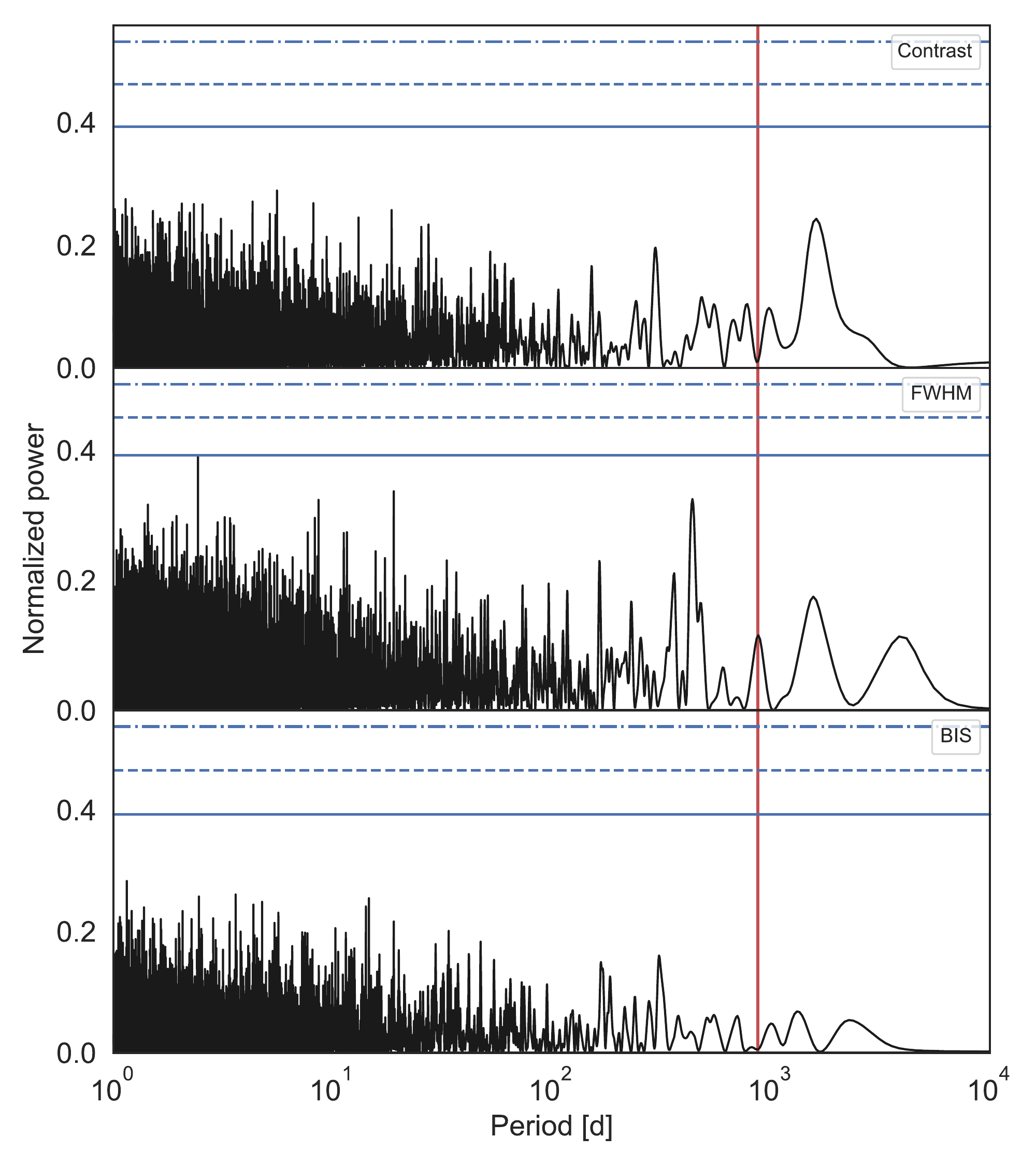}\\
        \caption{Periodogram of full width at half maximum (first panel), Bisector inverse span (second panel), and contrast (third panel) for HD22532. The red vertical line represents the fit period in the radial velocity at 872.6\,days. Horizontal lines, from bottom to top, are the FAP levels at 10\%, 1\%, and 0.1\%, respectively.}
        \label{fig:hd22532_perio_indic}
\end{figure}

\begin{figure}[!ht]
        \centering
        \adjincludegraphics[width=\columnwidth, trim={0 0 0 0},clip]{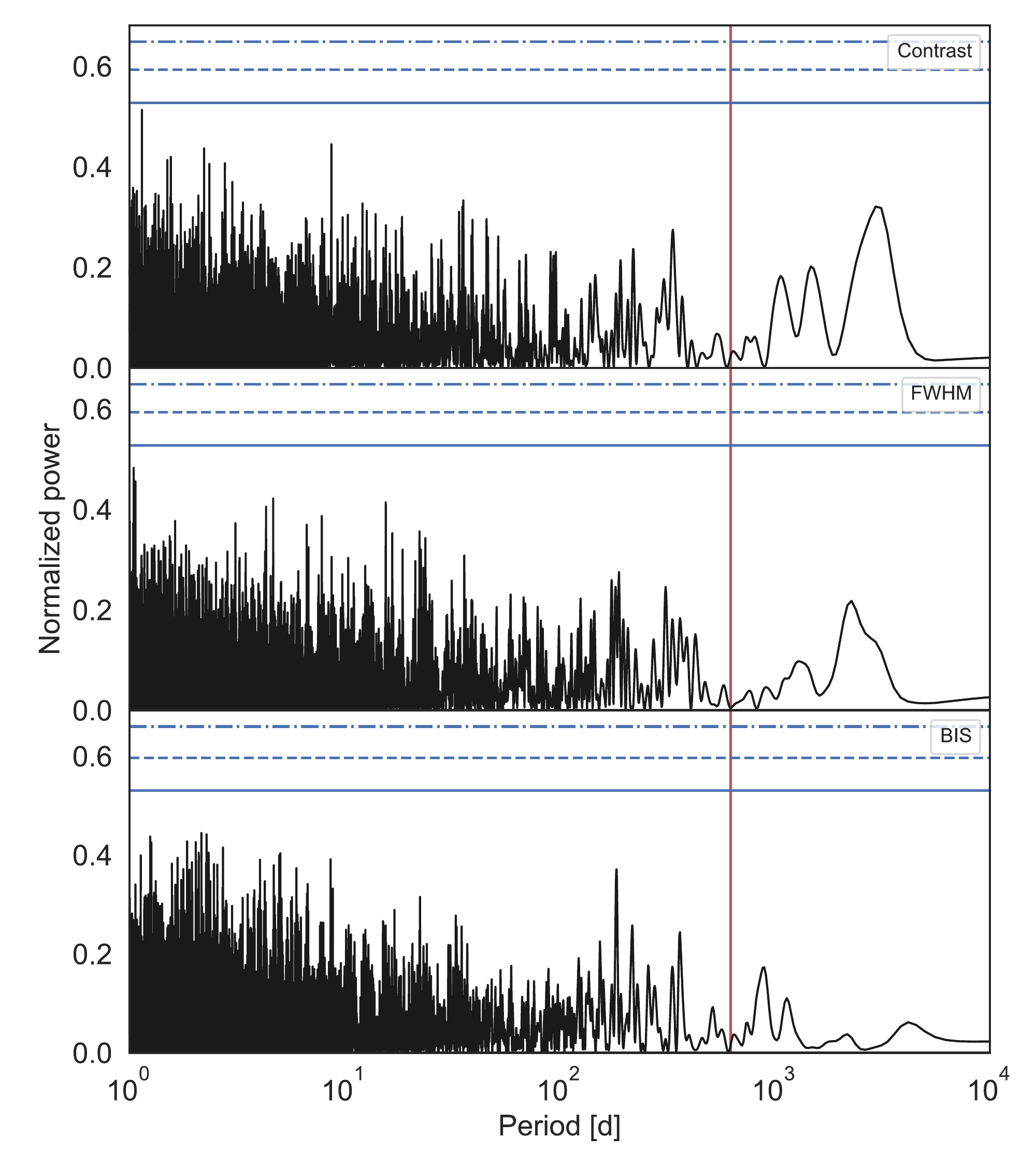}\\
        \caption{Same as \autoref{fig:hd22532_perio_indic}, but for HD\,64121. The period of the best solution is 623.0\,days.}
        \label{fig:hd64121_perio_indic}
\end{figure}

\begin{figure}[!ht]
        \centering
        \adjincludegraphics[width=\columnwidth, trim={0 0 0 0},clip]{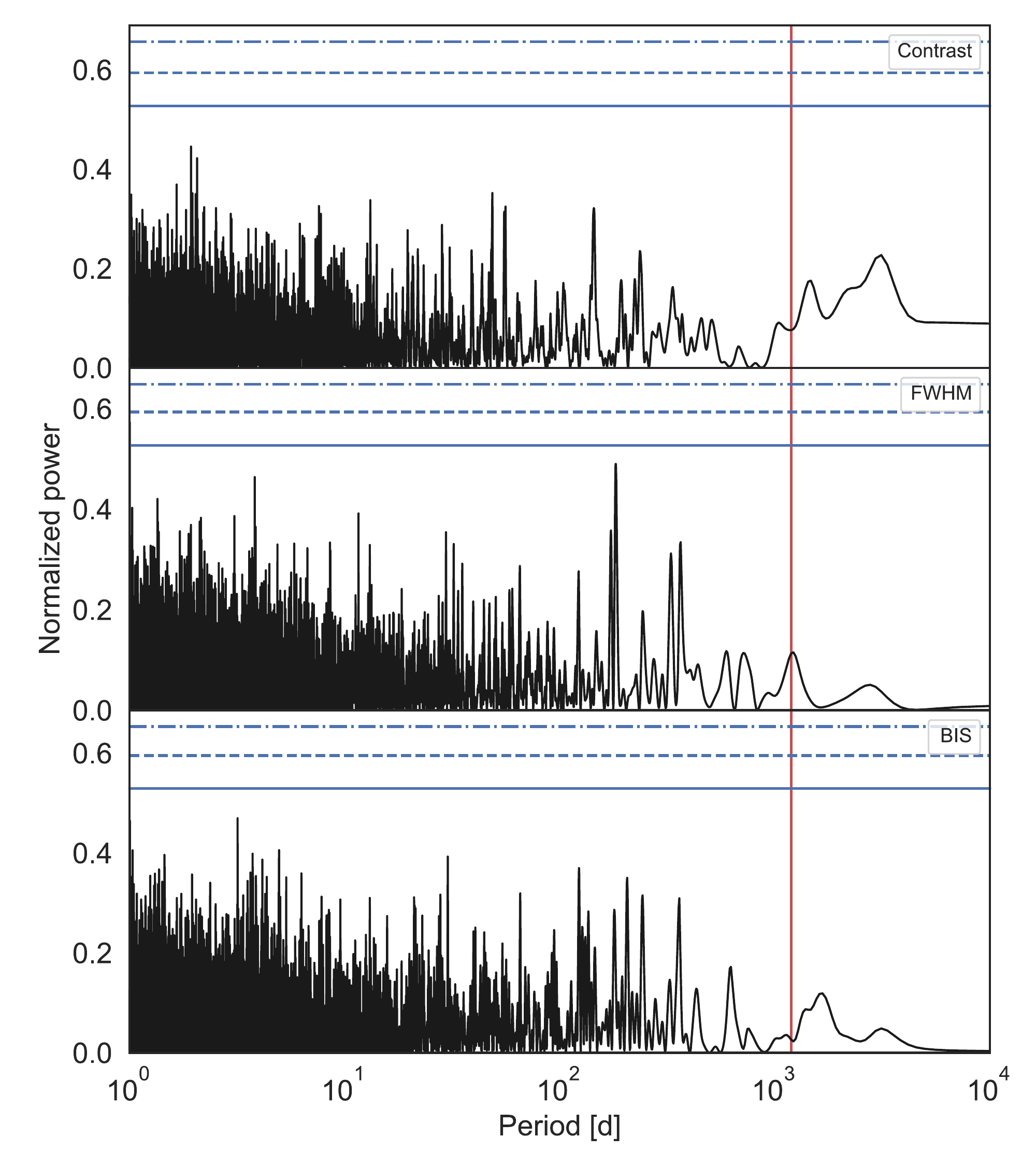}\\
        \caption{Same as \autoref{fig:hd22532_perio_indic}, but for HD\,69123. The period of the best solution is 1193.3\,days.}
        \label{fig:hd69123_perio_indic}
\end{figure}

\end{appendix}

\end{document}